\documentclass[12pt,a4]{article}

\usepackage[dvips]{graphicx}

\usepackage{amsmath,amssymb}
\usepackage{amsfonts}
\usepackage{bm}

\newcommand{\tr}{{\rm tr}}

\newcommand{\ol}{\overline}

\newcommand{\wt}{\widetilde}
\newcommand{\wh}{\widehat}

\newcommand{\rap}[2]
{\setbox1=\hbox{#1}%
\setbox2=\hbox to\wd1{\hss #2\hss}%
\mbox{\rlap{\box1}\box2}}

\newcommand{\sla}[1]{\rap{$#1$}{$\backslash$}}

\def\Det{\mathop{\rm Det}\nolimits}
\def\det{\mathop{\rm det}\nolimits}
\def\diag{\mathop{\rm diag}\nolimits}
\def\sign{\mathop{\rm sign}\nolimits}

\begin{document}
\begin{titlepage}
\title{
\begin{flushright}
\normalsize{TIT/HEP-613\\
Sep 2011}
\end{flushright}
       \vspace{2cm}
${\cal N}=2$ supersymmetric theories\\on squashed three-sphere
       \vspace{2cm}}
\author{
Yosuke Imamura\thanks{E-mail: \tt imamura@phys.titech.ac.jp}\quad
and\ \ Daisuke Yokoyama\thanks{E-mail: \tt d.yokoyama@th.phys.titech.ac.jp}
\\[30pt]
{\it Department of Physics, Tokyo Institute of Technology,}\\
{\it Tokyo 152-8551, Japan}
}
\date{}

\maketitle
\thispagestyle{empty}

\vspace{0cm}

\begin{abstract}
\normalsize
We investigate a squashing deformation of
3d ${\cal N}=2$ supersymmetric theories on three-sphere,
which have four supercharges.
The deformation preserves $SU(2)_L\times U(1)_r$ isometry
and all four supersymmetries.
We compute the partition function
and find non-trivial dependence on the squashing parameter.
We also consider the large $N$ limit of a certain class of
quiver gauge theories
which have free energy of order $N^{3/2}$,
and show that the free energy on the squashed sphere differs from that on round sphere by
a certain factor depending only on the squashing parameter.
\end{abstract}

\end{titlepage}

\tableofcontents

\section{Introduction}
Recently, exactly calculable quantities in gauge theories
in various dimensions attract great interest.
They provide strong evidences for the duality
among field theories and the gauge/gravity correspondence.
They are also useful to study relations among theories in different dimensions.

In the case of three dimensional theories,
the superconformal index\cite{Bhattacharya:2008zy,Kim:2009wb,Imamura:2011su}
and ${\bf S}^3$ partition function\cite{Kapustin:2009kz,Jafferis:2010un,Hama:2010av}
are such calculable quantities.
The ${\bf S}^3$ partition function was first computed
for superconformal theories without anomalous dimensions
in \cite{Kapustin:2009kz} by using localization.
It was later extended to theories with ${\cal N}=2$
superconformal theories\cite{Jafferis:2010un,Hama:2010av}.
It is used to check dualities among 3d field theories\cite{Kapustin:2010mh,Kapustin:2011gh,Willett:2011gp,Dolan:2011rp,Jafferis:2011ns,Benini:2011mf,Hwang:2011ht} and gauge/gravity correspondence\cite{Drukker:2010nc,Herzog:2010hf,Martelli:2011qj,Cheon:2011vi,Jafferis:2011zi}.
The superconformal index is also used to check these dualities\cite{Imamura:2011uj,Cheon:2011th,Krattenthaler:2011da,Hwang:2011qt,Hwang:2011ht}.

The superconformal index depends on chemical potentials
associated with global symmetries of the theory.
Similarly, the ${\bf S}^3$ partition function
is a function of the deformation parameters of the theory.
Studying the dependence of the partition function
on deformation parameters is important
because the more deformation parameters provided, the finer the information of the theory.
The partition function is given in the integral form\cite{Kapustin:2009kz}
\begin{equation}
Z=\int d\sigma_0 e^{-S^{\rm cl}(\sigma_0)}Z^{\rm 1-loop}(\sigma_0),
\label{zintf}
\end{equation}
where $\sigma_0$ is the expectation value of the adjoint scalar fields
in vector multiplets, which parametrizes the flat directions.
The integral is performed over the Cartan algebra of the gauge group.
In the most general form of the partition function,
it depends on
the Weyl weight of chiral multiplets,
real mass parameters, FI parameters,
Chern-Simons levels, and a squashing parameter of the ${\bf S}^3$.
FI parameters and Chern-Simons levels
enter in the partition function through the classical action $S^{\rm cl}(\sigma_0)$.
The real mass parameters enters through the one-loop determinant
of chiral multiplets, and can be introduced as expectation values
of scalar fields in external vector multiplets
coupling to flavor currents.
The partition function for general Weyl weight assignments is computed
in \cite{Jafferis:2010un,Hama:2010av}.
In 3d ${\cal N}=2$ superconformal field theories the Weyl weight $\Delta$ of a chiral multiplet
is the same as the superconformal R charge of the chiral multiplet.
Theories we consider in this paper are not always conformal.
When we consider non-conformal ${\cal N}=2$ supersymmetric theories on ${\bf S}^3$,
the Weyl weight should be regarded as a parameter appearing in the supersymmetry transformation laws
of chiral multiplets.

The squashing parameter was first introduced in \cite{Hama:2011ea}.
They consider two kinds of squashed ${\bf S}^3$.
The first one
is the squashed sphere with the metric
\begin{equation}
ds^2=r^2\left[(\mu^1)^2+(\mu^2)^2+\frac{1}{v^2}(\mu^3)^2\right].
\label{sqmetric}
\end{equation}
$\mu^a$ ($a=1,2,3$) are the
left-invariant differentials on ${\bf S}^3\sim SU(2)$ defined by
\begin{equation}
2\mu^aT_a=g^{-1}dg,\quad
g\in SU(2).
\label{lioneform}
\end{equation}
We use anti-Hermitian $SU(2)$ generators $T_a$ ($a=1,2,3$)
satisfying the commutation relations
\begin{equation}
[T_a,T_b]=-\epsilon_{abc}T_c,\quad
\epsilon_{123}=1.
\label{su2comm}
\end{equation}
We define symmetries $SU(2)_L$ and $SU(2)_R$ as left and right $SU(2)$ actions, respectively.
\begin{equation}
g\rightarrow g_Lgg_R,\quad
g_L\in SU(2)_L,\quad
g_R\in SU(2)_R.
\end{equation}
The parameter $v$ in the metric
(\ref{sqmetric}) is the squashing parameter.%
\footnote{The parameters $v$ and $r$ used in this paper are
related to ones in \cite{Hama:2011ea} by
$\ell=r$,
$\wt\ell=r/v$,
and $f=rv$.
$u$ is related to $s$ in \cite{Imamura:2011uw} by $u=is$.}
For later convenience we also define $u$ by
\begin{equation}
v^2=1+u^2.
\end{equation}
The round sphere corresponds to $v=1$ and $u=0$.
The differentials $\mu^a$ are invariant under $SU(2)_L$, while
they are transformed as a triplet under $SU(2)_R$.
Therefore, when $v\neq1$, the metric (\ref{sqmetric}) breaks $SU(2)_R$ to
its $U(1)$ subgroup, which is denoted by $U(1)_r$.

${\cal N}=2$ superconformal theories on round ${\bf S}^3$ have
eight supersymmetries,
and the squashing breaks them.
\cite{Hama:2011ea}
shows that it is possible to recover $1/4$ of them
(two supersymmetries) by turning on a Wilson line for the R-symmetry.
It is important that the recovered supersymmetries are $SU(2)_L$ singlets.
They computed the ${\bf S}^3$ partition function for such theories
with the expectation that they may obtain
a result depending on the squashing parameter in a non-trivial way.
The result was rather disappointing.
It was turned out that the partition function
is identical to that
on the round sphere up to some variable changes.

Having obtained this result,
the authors of \cite{Hama:2011ea} moved on to study another model
in which both $SU(2)_L$ and $SU(2)_R$ are broken.
This squashed sphere is often called ``ellipsoid.''
They again turn on an R-symmetry Wilson line to recover $1/4$ supersymmetry,
and compute the partition function.
This time they obtained
the $1$-loop partition function
\begin{align}
Z^{\rm 1-loop}
=\frac{\prod_{\alpha\in\Delta}s_b(x_0(\alpha(\sigma_0)))}
{\prod_I\prod_{\rho\in {\cal R}_I}s_b(x_{\Delta_I}(\rho(\sigma_0)))},
\label{zvc}
\end{align}
with the parameter $b$ depending on the squashing parameter of
the ellipsoid in a certain way.
The numerator is the contribution of vector multiplets,
and $\alpha$ runs over all roots of the gauge algebra.
The denominator contains the contribution of chiral multiplets.
$I$ labels chiral multiplets, and ${\cal R}_I$ and $\Delta_I$ are
the gauge representation and the Weyl weight, respectively,
of a chiral multiplet $I$.
$\rho$ runs over weights in the representation ${\cal R}_I$.
$s_b(x)$ is the double sine function defined by
\begin{equation}
s_b(x)
=
\prod_{p,q=0}^\infty
\frac{pb+qb^{-1}+\frac{Q}{2}-ix}
{pb+qb^{-1}+\frac{Q}{2}+ix},\quad
Q=b+\frac{1}{b}.
\end{equation}
$x_\Delta$ and $x_0$ are defined by
\begin{equation}
x_0(\alpha(\sigma_0))=\sqrt{\frac{Q}{2}}r\alpha(\sigma_0)-\frac{iQ}{2},\quad
x_\Delta(\rho(\sigma_0))=\sqrt{\frac{Q}{2}}r\rho(\sigma_0)-\frac{iQ}{2}(1-\Delta).
\end{equation}

To understand the independence of the partition function on the
squashing parameter of the $SU(2)_L\times U(1)_r$ symmetric squashing in \cite{Hama:2011ea},
let us consider which modes of fields contribute to the partition function.
Let us focus on a chiral multiplet.
Its contribution to the $1$-loop partition function
is given by
\begin{equation}
Z^{\rm 1-loop}=\frac{\Det {\cal D}_F}{\Det{\cal D}_B},
\label{ch1lop}
\end{equation}
where
${\cal D}_B$ and ${\cal D}_F$ are certain differential operators
appearing in the scalar and fermion actions.
Their determinants are the products of
eigenvalues of the differential operators.
A complex scalar field on ${\bf S}^3$ can be expanded by scalar spherical harmonics,
which belong to the $SU(2)_L\times SU(2)_R$ representation
\begin{equation}
\left(\bigoplus_{j=0}^\infty (j,j)_B\right)
\oplus 
\left(\bigoplus_{j=0}^\infty (j,j)_B\right).
\end{equation}
We use subscripts `$B$' and `$F$' to indicate the statistics of modes.
Roughly speaking, the two summations correspond to particles and anti-particles.
Similarly, a spinor field is expanded as
\begin{equation}
\left(\bigoplus_{j=0}^\infty (j+1/2,j)_F\right)
\oplus 
\left(\bigoplus_{j=0}^\infty (j,j+1/2)_F\right).
\end{equation}
Because of supersymmetry, the majority of these modes are paired between bosons and fermions,
and their contribution to the partition function
(\ref{ch1lop})
cancel each other.
If there exists an $SU(2)_L$ singlet supercharge, which is actually the case
in the $SU(2)_L\times U(1)_r$ symmetric squashing in \cite{Hama:2011ea},
the cancellation occurs between modes with the same $SU(2)_L$ quantum numbers:
\begin{equation}
(j,j)_B\leftrightarrow(j,j-1/2)_F,\quad
\mbox{or}\quad
(j,j)_B\leftrightarrow(j,j+1/2)_F.
\label{pairs}
\end{equation}
In the first pair in (\ref{pairs})
the number of bosonic modes in $(j,j)$ is larger than that of the fermionic modes in $(j,j-1/2)$.
After the cancellation, only the bosonic modes with the highest or lowest $SU(2)_R$ weight
survive and
contribute to the $1$-loop partition function
(\ref{ch1lop}).
Similarly, in the second pair in (\ref{pairs}),
only the fermionic modes with the highest or lowest $SU(2)_R$ weight
contribute to the partition function (\ref{ch1lop}).
Thus, even if the $SU(2)_R$ symmetry is broken and the degeneracy in
each $SU(2)_R$ multiplet is lost, it does not affect the structure of the partition function.
This is also the case for vector multiplets.%
\footnote{We would like to thank K.~Hosomichi
for pointing out the importance of the symmetry breaking.}

From the arguments above, we notice that
if we can realize squashing without
$SU(2)_L$ singlet supercharges
we may obtain the partition function depending
on the squashing parameter in a non-trivial way
even if the ${\bf S}^3$ is $SU(2)_L\times U(1)_r$-symmetric.
To study such theories is a main purpose of this paper.
One way to construct such theories is to compactify 4d theories by ${\bf S}^1$.
Let us consider a 4d ${\cal N}=1$ superconformal theory on ${\bf S}^3\times{\bf R}$.
The isometry of this background is $SU(2)_L\times SU(2)_R\times {\bf R}$.
The theory has eight supersymmetries,
and it is possible to compactify ${\bf R}$ to ${\bf S}^1$
with preserving four supersymmetries
belonging to $SU(2)_L$ doublets\cite{Imamura:2011uw}.
Through this compactification, we can relate
the ${\bf S}^3$ partition function to
the 4d superconformal index\cite{Dolan:2011rp,Gadde:2011ia,Imamura:2011uw}.
It is pointed out in \cite{Gadde:2011ia} that if we turn on the $SU(2)_R$ Wilson line, we can reproduce $1$-loop partition function (\ref{zvc}) with $b\neq1$
from 4d superconformal index.
The 3d theory obtained by such a compactification is a theory in squashed ${\bf S}^3$ with $SU(2)_L\times U(1)_r$ isometry,
and is different from the theories studied in \cite{Hama:2011ea}.
We give the supersymmetry transformation laws
and Lagrangians on the squashed sphere,
and compute the partition function.

Furthermore, we study the free energy of large $N$ gauge theories.
AdS/CFT correspondence relates 3d quiver gauge theories
on round ${\bf S}^3$ to M-theory in the background AdS$_4\times M_7$
with various compact 7-manifolds $M_7$.
The analysis on the gravity side claims that
the free energy is proportional to $N^{3/2}$,
and this has been confirmed on the gauge theory side
for a large class of theories\cite{Drukker:2010nc,Herzog:2010hf,Jafferis:2011zi}
when the background is round ${\bf S}^3$.
We extend the analysis on the gauge theory side
to the squashed sphere,
and determine the dependence of free energy on the squashing parameter.

This paper is organized as follows.
In Section \ref{theory.sec}, we give the supersymmetry transformation laws and
supersymmetric Lagrangians without derivations.
In Section \ref{loop.sec} we compute the $1$-loop partition function and
obtain (\ref{zvc})
with the parameters
\begin{equation}
b=\frac{1+iu}{v},\quad
x_0(\alpha(\sigma_0))=\frac{r\alpha(\sigma_0)-i}{v},\quad
x_\Delta(\rho(\sigma_0))=\frac{r\rho(\sigma_0)-i(1-\Delta)}{v}.
\label{bxx0}
\end{equation}
In Section \ref{4d3d.sec}, we explain how we can derive
the transformation laws and Lagrangians
given in Section \ref{theory.sec}
by the dimensional reduction from 4d theory.
In Section \ref{largen.sec} we study the free energy of large $N$ quiver gauge theories
which are expected to have M-theory duals.
Section \ref{conc.sec} is devoted to our conclusions.

Before ending this section, we summarize our conventions and notations.
We use the $SU(2)_L$-invariant local frame on the squashed sphere
with the vielbein
\begin{equation}
e^{\wh1}=r\mu^1,\quad
e^{\wh2}=r\mu^2,\quad
e^{\wh3}=\frac{r}{v}\mu^3.
\label{squasheds3e}
\end{equation}
We use Roman characters $k,l,m,n,\ldots,=1,2,3$ for 3d tangent indices,
and hatted characters $\wh k,\wh l,\wh m,\wh n,\ldots,=\wh 1,\wh 2,\wh3$ for
local indices.
Three-dimensional spinors have two components,
and Dirac's matrices $\gamma^{\wh m}$ are $2\times2$ matrices.
We use $\gamma^{\wh m}=\sigma_m$, where
$\sigma_m$ are the Pauli's matrices
\begin{equation}
\sigma_1
=\left(\begin{array}{cc} 0 & 1 \\
1 & 0 \end{array}\right),\quad
\sigma_2
=\left(\begin{array}{cc} 0 & -i \\
i & 0 \end{array}\right),\quad
\sigma_3
=\left(\begin{array}{cc} 1 & 0 \\
0 & -1 \end{array}\right).
\end{equation}
The Levi-Civita tensor is defined by $\epsilon_{\wh 1\wh2\wh 3}=1$.
We use spinors without and with bars, which are transformed in the same way
under 3d rotations.
Two kinds of spinors
originate from left-handed and right-handed spinors when we derive the
theory from 4d theory by the dimensional reduction in Section
\ref{4d3d.sec}.

\section{${\cal N}=2$ supersymmetry on the squashed sphere}\label{theory.sec}
\subsection{Transformation laws}
${\cal N}=2$ superconformal theories
on round ${\bf S}^3$ have eight supercharges.
If we turn on real mass parameters, half of the supersymmetries are
broken, and we call the unbroken part ${\cal N}=2$ supersymmetry.
It is possible to squash the ${\bf S}^3$ in such a way that the ${\cal N}=2$
supersymmetry is preserved.
Killing spinors $\epsilon$ and $\ol\epsilon$ for the four supersymmetries
satisfy
the Killing equations
\begin{align}
D_m\epsilon
&=-\frac{i}{2vr}\gamma_m\epsilon
+\frac{u}{vr}f^n\gamma_{mn}\epsilon,\nonumber\\
D_m\ol\epsilon
&=-\frac{i}{2vr}\gamma_m\ol\epsilon
-\frac{u}{vr}f^n\gamma_{mn}\ol\epsilon,
\label{3dkilling}
\end{align}
where we define the vector field
\begin{equation}
f^m=e^m_{\wh 3}.
\end{equation}
This vector field generates $U(1)_r$ isometry.
Each of the differential equations in (\ref{3dkilling})
has two linearly independent solutions which form an $SU(2)_L$ doublet.
An explicit form of the solutions are
\begin{equation}
\epsilon=e^{-\theta T_3}g^{-1}\epsilon_0,\quad
\ol\epsilon=e^{\theta T_3}g^{-1}\ol\epsilon_0,
\end{equation}
where $\epsilon_0$ and $\ol\epsilon_0$ are arbitrary constant spinors,
and $\theta$ is the angle defined by $e^{i\theta}=(1+iu)/v$.

Supersymmetry transformation laws for component fields of vector multiplets
are
\begin{align}
\delta A_m&=
i(\epsilon\gamma_m\ol\lambda)
-i(\ol\epsilon\gamma_m\lambda)
+uf_m(\epsilon\ol\lambda)
+uf_m(\ol\epsilon\lambda),\nonumber\\
\delta\sigma&=v(\epsilon\ol\lambda)+v(\ol\epsilon\lambda),\nonumber\\
\delta\lambda&=-{\cal F}_{\wh m}^{(+)}\gamma_{\wh m}\epsilon+D\epsilon,\nonumber\\
\delta\ol\lambda&={\cal F}_{\wh m}^{(-)}\gamma_{\wh m}\ol\epsilon+D\ol\epsilon,\nonumber\\
\delta D&=
-(\epsilon\gamma^m D_m\ol\lambda)
+\frac{i}{2vr}(\epsilon\ol\lambda)
+\frac{1}{v}(\epsilon(1-iu\sla f)[\sigma,\ol\lambda])
\nonumber\\&\quad
-(\ol\epsilon\gamma^m D_m\lambda)
+\frac{i}{2vr}(\ol\epsilon\lambda)
-\frac{1}{v}(\ol\epsilon(1+iu\sla f)[\sigma,\lambda]),
\end{align}
where $\sla f=f^m\gamma_m$ and ${\cal F}_{\wh m}^{(\pm)}$ are defined by
\begin{equation}
{\cal F}_{\wh m}^{(\pm)}
=\frac{1}{2}\epsilon_{\wh m\wh p\wh q}F_{\wh p\wh q}
 +\frac{u}{v}f^{\wh p}\epsilon_{\wh m\wh p\wh n}D_{\wh n}\sigma
 \pm \frac{1}{v}D_{\wh m}\sigma.
\label{vin3d}
\end{equation}
Transformation laws for component fields in a chiral multiplet with Weyl weight $\Delta$ are
\begin{align}
\delta \phi&=\sqrt2(\epsilon\psi),\nonumber\\
\delta \phi^\dagger&=\sqrt2(\ol\epsilon\ol\psi),\nonumber\\
\delta\psi&=
-\sqrt{2}\gamma^m\ol\epsilon D_m\phi
+\frac{\sqrt2}{v}(1-iu\sla f)\ol\epsilon \sigma\phi
+\sqrt{2}\epsilon F
+\frac{\sqrt2\Delta i}{vr}(1-iu\sla f)\ol\epsilon\phi
,\nonumber\\
\delta\ol\psi&=
-\sqrt{2}\gamma^m\epsilon D_m\phi^\dagger
+\frac{\sqrt{2}}{v} (1+iu\sla f)\epsilon \phi^\dagger\sigma
+\sqrt{2}\ol\epsilon F^\dagger
+\frac{\sqrt2\Delta i}{vr}(1+iu\sla f)\epsilon\phi^\dagger
,\nonumber\\
\delta F&=
-\sqrt{2}D_m(\ol\epsilon\gamma^m\psi)
-\frac{\sqrt2(\Delta-2)i}{vr}(\ol\epsilon(1+iu\sla f)\psi)
\nonumber\\&\hspace{10em}
-\frac{\sqrt2}{v}(\ol\epsilon(1+iu\sla f)\sigma\psi)
-2(\ol\epsilon\ol\lambda)\phi,
\nonumber\\
\delta F^\dagger&=
-\sqrt{2}D_m(\epsilon\gamma^m\ol\psi)
-\frac{\sqrt2(\Delta-2)i}{vr}(\epsilon(1-iu\sla f)\ol\psi)
\nonumber\\&\hspace{10em}
-\frac{\sqrt2}{v}(\epsilon(1-iu\sla f)\ol\psi)\sigma
-2\phi^\dagger(\epsilon\lambda).
\end{align}

The commutation relation of the two transformations $\delta(\epsilon,\ol\epsilon)$ and
$\delta(\epsilon',\ol\epsilon')$ is
\begin{equation}
[\delta(\epsilon,\ol\epsilon),\delta(\epsilon',\ol\epsilon')]
=2{\cal L}_{l'}+2\alpha\left(-i\sigma+\frac{R}{r}\right).
\label{comm}
\end{equation}
$R$ is the R charge, and $\sigma$ should be understood as
the gauge transformation with parameter $\sigma$.
$l'$ and $\alpha$ are bilinear of the transformation parameters
\begin{equation}
l'^m=(\ol\epsilon\gamma^m\epsilon')+(\epsilon\gamma^m\ol\epsilon'),\quad
\alpha
=\frac{i}{v}\ol\epsilon(1+iu\sla f)\epsilon'
-\frac{i}{v}\epsilon(1-iu\sla f)\ol\epsilon',
\end{equation}
and ${\cal L}_v$ is the Lie derivative associated with a vector field $v$.
It is easily shown by the Killing equations (\ref{3dkilling}) that $l'^m$ is a Killing vector
and $\alpha$ is a constant on the squashed sphere.
$l'^m$ can be divided into a $SU(2)_L$ part $l^m$ and $U(1)_r$ part proportional to
$f^m$:
\begin{equation}
l'^m=l^m-\frac{u}{v}\alpha f^m.
\label{lpla}
\end{equation}
The right hand side
in (\ref{comm}) contains
generators of $SU(2)_L$, $U(1)_r$, and $U(1)_R$.
$U(1)_r$ does not rotate the supercharges, and thus is the center of the algebra.
Therefore, the supersymmetry algebra on the squashed sphere is $SU(2|1)\ltimes U(1)_r$, a central extension
of $SU(2|1)$.
If we regard the 3d theory as an ${\bf S}^1$ compactification of a
4d theory,
$\alpha$ can be regarded as the parameter of a shift along the $4$-th direction.
If we substitute
(\ref{lpla}) into
(\ref{comm}),
we have $U(1)_r$ transformation with $\alpha$ in the coefficient.
This implies the existence of non-vanishing graviphoton background field.
From the 4d perspective, a graviphoton field is, roughly speaking, identified with the
non-diagonal components $g_{m4}$ of the metric.
When the background graviphoton field is non-vanishing,
the compactified direction $x^4$ is tilted, and shift along $x^4$
generates a shift in 3d proportional to the graviphoton potential
field when it is projected onto 3d.
(\ref{lpla}) implies that the graviphoton field in our background is given by
\begin{equation}
V^m=\frac{u}{v}f^m.
\label{graviphoton}
\end{equation}
We will see
in Section \ref{4d3d.sec} that the graviphoton field (\ref{graviphoton})
is indeed arises in the compactification.

\subsection{Actions}
The supersymmetric kinetic Lagrangian for vector multiplet is
\begin{equation}
{\cal L}_{\rm YM}
={\cal L}_{\cal A}+{\cal L}_\lambda-\frac{1}{2}\tr D^2,
\label{lym}
\end{equation}
where ${\cal L}_{\cal A}$ and ${\cal L}_\lambda$ are bosonic
and fermionic terms given by
\begin{align}
{\cal L}_{\cal A}
&=\frac{1}{2}\tr({\cal F}_{\wh m}^{(-)}{\cal F}_{\wh m}^{(-)}),
\nonumber\\
{\cal L}_\lambda
&=
\tr\left[-\ol\lambda\gamma^m D_m\lambda
+\frac{i}{2vr}\ol\lambda\lambda
-\frac{1}{v}\ol\lambda(1+iu\sla f)[\sigma,\lambda]\right].
\label{lall}
\end{align}
$\tr$ is a positive definite gauge invariant inner product of the gauge algebra.

The supersymmetric kinetic Lagrangian for chiral multiplet with Weyl weight $\Delta$ is
\begin{align}
{\cal L}_{\rm chiral}
&={\cal L}_\phi+{\cal L}_\psi-F^\dagger F,
\label{4dlag}
\end{align}
where ${\cal L}_\phi$ and ${\cal L}_\psi$ are given by
\begin{align}
{\cal L}_\phi
&=-\phi^\dagger D_m D^m\phi
+\phi^\dagger \sigma\sigma\phi
   +\phi^\dagger D\phi
   -\frac{\Delta^2-2\Delta}{r^2}\phi^\dagger\phi
+\frac{2i(\Delta-1)}{r}\phi^\dagger \sigma\phi
\nonumber\\&
\quad
+\frac{u}{v}f^m\left[
-i\phi^\dagger \sigma D_m\phi
-i\phi^\dagger D_m( \sigma\phi)
+\frac{2(\Delta-1)}{r}\phi^\dagger D_m\phi\right],
\nonumber\\
{\cal L}_\psi
&=
-(\ol\psi\gamma^m D_m\psi)
+\frac{i}{2vr}(\ol\psi\psi)
-\left(\ol\psi\frac{i(\Delta-ir\sigma)}{vr}(1+iu\sla f)\psi\right)
\nonumber\\&
\quad
   -\sqrt2\phi^\dagger(\lambda\psi)
   -\sqrt2(\ol\psi\ol\lambda)\phi.
\label{3daction}
\end{align}

Let $\ol\epsilon_1$ and $\ol\epsilon_2$ be two independent
solutions of the second equation in (\ref{3dkilling}).
The kinetic Lagrangians (\ref{lym}) and (\ref{4dlag})
can be obtained from
\begin{equation}
(\ol\epsilon_1\ol\epsilon_2)
{\cal L}_{\rm YM}=
-\frac{1}{4}\delta(\ol\epsilon_1)
\delta(\ol\epsilon_2)\tr(\ol\lambda\ol\lambda),\quad
(\ol\epsilon_1\ol\epsilon_2)
{\cal L}_{\rm chiral}=-\frac{1}{2}
\delta(\ol\epsilon_1)
\delta(\ol\epsilon_2)(\phi^\dagger F).
\end{equation}
Because $\delta(\ol\epsilon_1)$ and $\delta(\ol\epsilon_2)$ commute
with each other
the right hand side
of these equations contains the parameters $\ol\epsilon_1$ and $\ol\epsilon_2$
only through the scalar product $(\ol\epsilon_1\ol\epsilon_2)$, and
these equations consistently define the Lagrangians ${\cal L}_{\rm YM}$ and ${\cal L}_{\rm chiral}$.
These Lagrangians do not depend on the choice of two independent
Killing spinors $\ol\epsilon_1$ and $\ol\epsilon_2$,
and they are exact with respect to $\delta(\ol\epsilon)$ for any $\ol\epsilon$
satisfying (\ref{3dkilling}).

The supersymmetric completion of the Chern-Simons term and the FI term
are
\begin{align}
{\cal L}_{\rm CS}&=\tr_{\rm CS}\bigg[
\frac{i}{2}\epsilon^{mnp}\left(A_m\partial_n A_p-\frac{2i}{3}A_m A_n A_p\right)
\nonumber\\&\hspace{7em}
+(\ol\lambda\lambda)
-\frac{1}{v}D\sigma
+\frac{i}{vr}\sigma^2
-\frac{iu}{2v}\sigma\epsilon^{mnp}f_m F_{np}
\bigg]
,\nonumber\\
{\cal L}_{\rm FI}&=-\tr_{\rm FI}\left[D-\frac{2i}{r}\sigma+\frac{2ui}{vr}f^mA_m\right],
\label{csandfi}
\end{align}
where $\tr_{\rm CS}$ is a gauge invariant inner product of Lie algebra,
which does not have to be positive definite,
and $\tr_{\rm FI}$ is a gauge invariant linear map from the gauge algebra
to ${\bf R}$.

In addition to these, the $F$-components of gauge invariant chiral multiplets of
weight $\Delta=2$ are supersymmetry invariant up to total derivatives.
Such terms, however, do not affect the partition function.

\section{Partition function}\label{loop.sec}
In this section we compute the partition function of a theory
on the squashed ${\bf S}^3$.
Because of the $\delta(\ol\epsilon)$-exactness of the kinetic Lagrangians
${\cal L}_{\rm YM}$ and ${\cal L}_{\rm chiral}$,
we can send the coefficients of these Lagrangians
to infinity without changing the partition function.
The theory becomes free in this limit,
and we can perform the path integral
to obtain the expression (\ref{zintf}) of the partition function.
\subsection{Mode expansion on squashed ${\bf S}^3$}
Let $\Phi(g)$ be a spin $s$ field on the squashed sphere.
We expand it by the spin basis $|s,s_z\rangle$ ($s_z=-s,-s+1,\ldots,s$)
\begin{equation}
\Phi(g)=\sum_{s_z=-s}^s\Phi_{s_z}(g)|s,s_z\rangle.
\end{equation}
Because we are using the $SU(2)_L$-invariant frame, $|s,s_z\rangle$ are transformed as the $(0,s)$ representation
of $SU(2)_L\times SU(2)_R$.
$\Phi_{s_z}(g)$ for each $s_z$ is a scalar function on ${\bf S}^3$,
and can be expanded by the scalar spherical harmonics $Y^j_{m',m}(g)$
as
\begin{equation}
\Phi_{s_z}(g)=\sum_{j,m',m}\Phi_{s_z,m',m}^jY^j_{m',m}(g).
\end{equation}
The harmonics $Y^j_{m',m}$
belong to the $(j,j)$ representation of $SU(2)_L\times SU(2)_R$.
$j$ is the common azimuthal quantum number for both $SU(2)_L$ and $SU(2)_R$,
and $m'$ and $m$ are magnetic quantum numbers for $SU(2)_L$ and $SU(2)_R$,
respectively.
They take values
\begin{align}
j&=0,\frac{1}{2},1,\ldots,\nonumber\\
m&=-j,-j+1,\ldots,j-1,j,\nonumber\\
m'&=-j,-j+1,\ldots,j-1,j.
\label{jmmrange}
\end{align}
In the following, we use the ket notation for the harmonics $Y^j_{m',m}(g)$
\begin{equation}
|j,m',m\rangle
=Y^j_{m',m}(g).
\end{equation}
The expansion of the field $\Phi(g)$ is expressed as
\begin{equation}
\Phi(g)=\sum_{j,m',m,s_z}
\Phi_{s_z,m',m}^j|j,m',m\rangle\otimes |s,s_z\rangle.
\end{equation}

The covariant derivative on round ${\bf S}^3$
with the left-invariant frame
acts on the field $\Phi(g)$ as
\begin{equation}
D^{(0)}=\mu^a(2L_a+S_a),
\label{roundda}
\end{equation}
where $L_a$ and $S_a$ are $SU(2)$ generators.
$L_a$ are the $SU(2)_R$ orbital angular momenta acting on
the $SU(2)_R$ index $m$ of $|j,m',m\rangle$,
and $S_a$ are the spin operators acting on $|s,s_z\rangle$.
These operators are normalized so as to satisfy the commutation relation
(\ref{su2comm}).

The covariant derivative on the squashed sphere is obtained
from $D^{(0)}$ by replacing the spin connection on the round sphere,
$\omega_{(0)}^{\wh m\wh n}=\epsilon_{\wh m\wh n\wh p}\mu^p$,
by $\omega^{\wh m\wh n}$, the spin connection on the squashed sphere.
$\omega^{\wh m\wh n}$ and $\omega_{(0)}^{\wh m\wh n}$ are related by
\begin{align}
\omega^{\wh 1\wh 2}
&=\left(2-\frac{1}{v^2}\right)\mu^3
=\omega_{(0)}^{\wh 1\wh 2}+\left(1-\frac{1}{v^2}\right)\mu^3
,\nonumber\\
\omega^{\wh 2\wh 3}
&=\frac{1}{v}\mu^1
=\omega_{(0)}^{\wh 2\wh 3}+\left(\frac{1}{v}-1\right)\mu^1
,\nonumber\\
\omega^{\wh 3\wh 1}
&=\frac{1}{v}\mu^2
=\omega_{(0)}^{\wh 3\wh 1}+\left(\frac{1}{v}-1\right)\mu^2.
\label{sqspin}
\end{align}
Combining (\ref{roundda}) and (\ref{sqspin}),
we obtain
the following algebraic expression for the
covariant derivative on the squashed sphere.
\begin{equation}
D
=\mu^1\left(2L_{1}+\frac{1}{v}S_{1}\right)
+\mu^2\left(2L_{2}+\frac{1}{v}S_{2}\right)
+\mu^3\left[2L_{3}+\left(2-\frac{1}{v^2}\right)S_{3}\right].
\label{diffgen}
\end{equation}

The non-vanishing components of the spin $j$ representation matrices for generators $L_a$ are
\begin{align}
\langle j,m',m|L_3|j,m',m\rangle&=im,\nonumber\\
\langle j,m',m+\frac{1}{2}|L_{1+i2}|j,m',m-\frac{1}{2}\rangle&=i\sqrt{(j+\frac{1}{2})^2-m^2},\nonumber\\
\langle j,m',m-\frac{1}{2}|L_{1-i2}|j,m',m+\frac{1}{2}\rangle&=i\sqrt{(j+\frac{1}{2})^2-m^2},
\end{align}
where $L_{1\pm i2}\equiv L_1\pm iL_2$.
We also introduce $SU(2)_L$ generators $L'_a$.
The non-vanishing components of $L'_3$ are
\begin{equation}
\langle j,m',m|L'_3|j,m',m\rangle=im'.
\end{equation}

In the following subsections
we compute the determinant of certain differential operators appearing in the Lagrangians.
Because the squashed background preserves $SU(2)_L$ and $U(1)_r$,
the differential operators
commute with operators
$L'_aL'_a$,
$L_3'$,
and $L_3+S_3$.
Therefore,
we can compute the determinant
in each eigenspace defined by
\begin{equation}
L'_aL'_a=-j(j+1),\quad
L_3'=im',\quad
L_3+S_3=im.
\label{restricted}
\end{equation}
Because $L_a$ and $L_a'$ act on scalar spherical harmonics,
$L'_aL'_a=L_aL_a$ holds.
This restriction generically defines $2s+1$ dimensional vector space
spanned by
\begin{equation}
\{|j,m',m-s_z\rangle\otimes|s,s_z\rangle\}_{s_z=-s}^s,
\end{equation}
and the differential operator reduces to a $(2s+1)\times(2s+1)$
matrix on this subspace.
If $m$ is close to $\pm j$ and some $m-s_z$ are out of the allowed range in (\ref{jmmrange}),
special treatment is needed.

\subsection{Bosons in vector multiplets}
Because of the $\delta(\ol\epsilon)$-exactness of ${\cal L}_{\rm YM}$, we can add ${\cal L}_{\rm YM}$
to the Lagrangian of the theory
with an arbitrary coefficient without changing the partition function.
In the limit in which the coefficient goes to infinity,
the path integral for vector multiplet reduces to the Gaussian integral
around the saddle points.
Let us start with the bosonic part.
Saddle points are given by ${\cal F}_{\wh m}=D=0$.
This is the case iff
\begin{equation}
A_m=D=0,\quad
\sigma=\sigma_0,
\label{saddle}
\end{equation}
up to gauge transformations.
$\sigma_0$ is a constant expectation value of $\sigma$,
and we assume that it is diagonalized by gauge transformations.
At saddle points, the classical values of the Chern-Simons term and FI term in
(\ref{csandfi})
are
\begin{align}
S_{\rm CS}^{\rm cl}(\sigma_0)
&=\int d^3x\sqrt{g}{\cal L}_{\rm CS}^{\rm cl}(\sigma_0)
=\frac{2\pi^2 ir^2}{v^2}\tr_{\rm CS}(\sigma_0^2),\nonumber\\
S_{\rm FI}^{\rm cl}(\sigma_0)
&=\int d^3x\sqrt{g}{\cal L}_{\rm FI}^{\rm cl}(\sigma_0)
=\frac{4\pi^2 i r^2}{v}\tr_{\rm FI}(\sigma_0).
\label{csficl}
\end{align}
We define the fluctuation part of the scalar field
\begin{equation}
\varphi=\sigma-\sigma_0.
\end{equation}
The path integral of the auxiliary field $D$ gives constant, and we ignore its contribution.

All component fields in the vector multiplet belong to the adjoint representation
of the gauge group $G$, and have $\dim G$ components.
In the following, we focus on one component in each field that
satisfies $[\sigma_0,\Phi]=\alpha(\sigma_0)\Phi$.
To obtain the final expression, we need to take the product over all weights $\alpha$
in the adjoint representation.

To fix the gauge we introduce the gauge fixing function
\begin{equation}
f=D_{\wh m}A_{\wh m},
\label{gaugefix}
\end{equation}
and add the gauge fixing term
\begin{equation}
{\cal L}_{\rm GF}=\frac{1}{2}\tr f^2,
\end{equation}
to the Lagrangian.
We still have residual gauge symmetry with constant transformation parameters.
This residual symmetry is fixed by requiring the
constant mode of the scalar field $\sigma_0$ to be diagonal.
The Jacobian factor associated with
this gauge fixing is the Vandermonde determinant
\begin{equation}
\prod_{\alpha\in\Delta}\alpha(\sigma_0).
\label{vdm}
\end{equation}
We should include this factor to the result of the path integral below.

Let us define four-component field
${\cal A}=(A_{\wh1},A_{\wh2},A_{\wh3},\varphi)^T$.
In the following we ignore higher order terms with respect to the fluctuation fields.
The quadratic part of ${\cal L}_{\cal A}+{\cal L}_{\rm GF}$ with respect to
${\cal A}$ is
\begin{equation}
{\cal L}_{\cal A}+{\cal L}_{\rm GF}
=\frac{1}{2r^2}({\cal D}_{\cal A}{\cal A})^T
({\cal D}_{\cal A}{\cal A}),
\end{equation}
where the differential operator ${\cal D}_{\cal A}$
is defined by
\begin{equation}
{\cal D}_{\cal A}
\left(\begin{array}{c}
A_{\wh3} \\
A_{\wh1+i\wh2} \\
A_{\wh1-i\wh2} \\
\varphi
\end{array}\right)
=
r\left(\begin{array}{c}
{\cal F}_{\wh 3}^{(-)} \\
{\cal F}_{\wh 1+i\wh 2}^{(-)} \\
{\cal F}_{\wh 1-i\wh 2}^{(-)} \\
f
\end{array}\right).
\end{equation}
By using (\ref{diffgen})
with spin $1$ representation matrix
$(S_a)_{\wh b\wh c}=\epsilon_{abc}$,
we can rewrite the definition of ${\cal F}^{(-)}_{\wh m}$ in (\ref{vin3d})
in the algebraic form
\begin{align}
r{\cal F}_{\wh 3}^{(-)}
&=
\frac{2-ir\alpha(\sigma_0)}{v}A_{\wh 3}
-iL_{1-i2}A_{\wh 1+i\wh 2}
+iL_{1+i2}A_{\wh 1-i\wh 2}
-\frac{1}{v}2vL_{3}\varphi
,\nonumber\\
r{\cal F}_{\wh 1+i\wh 2}^{(-)}
&=
-2iL_{1+i2}A_{\wh 3}
+\left[2v(1+iL_{3})-\frac{1-iu}{v}ir\alpha(\sigma_0)\right] A_{\wh1+i\wh2}
-\frac{1-iu}{v}2L_{1+i2}\varphi
,\nonumber\\
r{\cal F}_{\wh 1-i\wh 2}^{(-)}
&=
2iL_{1-i2}A_{\wh 3}
+\left[2v(1-iL_{3})-\frac{1+iu}{v}ir\alpha(\sigma_0)\right] A_{\wh1-i\wh2}
-\frac{1+iu}{v}2L_{1-i2}\varphi
.
\end{align}
We also rewrite the gauge fixing function (\ref{gaugefix})
as
\begin{equation}
rf=
2vL_{3}A_{\wh 3}
+L_{1-i2}A_{\wh 1+i\wh 2}
+L_{1+i2}A_{\wh 1-i\wh 2}.
\label{algf}
\end{equation}
The algebraic form of ${\cal D}_{\cal A}$ is
\begin{equation}
{\cal D}_{\cal A}=\left(\begin{array}{cccc}
\frac{2-ir\alpha(\sigma_0)}{v} & -iL_{1-i2} & iL_{1+i2} & -2L_{3} \\
-2iL_{1+i2} & 
2v(1+iL_{3})-\frac{1-iu}{v}ir\alpha(\sigma_0) & 0 &
-\frac{1-iu}{v}2L_{1+i2} \\
2iL_{1-i2} & 0 &
2v(1-iL_{3})-\frac{1+iu}{v}ir\alpha(\sigma_0) &
-\frac{1+iu}{v}2L_{1-i2} \\
2vL_{3} & L_{1-i2} & L_{1+i2} & 0
\end{array}\right).
\label{opm}
\end{equation}
By restriction to the subspace defined by (\ref{restricted}),
the operator (\ref{opm}) becomes $4\times 4$ matrix
with each component being a complex number.
Its determinant is
\begin{equation}
\det {\cal D}_{\cal A}=\frac{4[j(j+1)+u^2m^2]}{v}(2j+2imu+ir\alpha(\sigma_0))(2j+2-2imu-ir\alpha(\sigma_0)).
\end{equation}
(We use ``$\det$'' for the determinant of the matrix defined in the subspace (\ref{restricted}),
and ``$\Det$'' for the functional determinant of differential operators.)
We need to divide this by the Jacobian factor associated with
the gauge fixing.
The algebraic form of the gauge transformation of ${\cal A}$ is
\begin{equation}
\delta{\cal A}=\left(\begin{array}{c}
\delta A_{\wh a}=D_{\wh a}\lambda \\
\delta\varphi=i[\lambda,\sigma_0]
\end{array}\right)
=\frac{1}{r}\left(\begin{array}{c}
2v L_{3} \\
2L_{1-i2} \\
2L_{1+i2} \\
-ir\alpha(\sigma_0)
\end{array}\right)\lambda.
\end{equation}
Substituting this into (\ref{algf}), we obtain
the Jacobian
\begin{equation}
r\frac{\delta f}{\delta\lambda}=-4[j(j+1)+u^2m^2].
\end{equation}
Therefore, the path integral of physical modes
in the
restricted vector space with the quantum numbers
in (\ref{restricted}) gives%
\footnote{We ignore constant factor $(-1/v)$.}
\begin{equation}
\det'{\cal D}_{\cal A}=\frac{\det{\cal D}_{\cal A}}{r\delta f/\delta\lambda}=(2j+2imu+ir\alpha(\sigma_0))(2j+2-2imu-ir\alpha(\sigma_0))
\label{detda}
\end{equation}
The two factors
in (\ref{detda})
correspond to the first two irreducible representation
in the decomposition
\begin{equation}
(j,j)\otimes (0,1)
=(j,j-1)\oplus(j,j+1)\oplus (j,j).
\label{vectormodes}
\end{equation}
The last representation corresponds to the gauge degrees of freedom.
By taking the product over quantum numbers $j$, $m$, and $m'$,
we obtain
\begin{equation}
\Det'{\cal D}_{\cal A}
=\prod_j
\prod_{|m|\leq j-1}(2j+2imu+ir\alpha(\sigma_0))^{2j+1}
\prod_{|m|\leq j+1}(2j+2-2imu-ir\alpha(\sigma_0))^{2j+1}.
\label{detpa0}
\end{equation}
Because the four supersymmetries are $SU(2)_R$ singlets,
the cancellation between bosons and fermions occurs
among the modes with the same $SU(2)_R$ quantum numbers.
For this reason, we shift the quantum number $j$ so that
the $SU(2)_R$ spins become $j$.
Namely, in the first factor in (\ref{detpa0}) we replace $j$ by
$j+1$, and in the second factor by $j-1$.
Correspondingly, the first two representations in (\ref{vectormodes})
become
\begin{equation}
(j+1,j)\oplus(j-1,j)
\label{vreps}
\end{equation}
After this shift we obtain
\begin{equation}
\Det'{\cal D}_{\cal A}
=\prod_j\prod_{|m|\leq j}(2j+2+2imu+ir\alpha(\sigma_0))^{2j+3}(2j-2imu-ir\alpha(\sigma_0))^{2j-1}.
\label{deta0}
\end{equation}

Up to now, we have not specified the region of the
spin $j$.
The product with respect to $j$ should be taken over the region
for which the spins in (\ref{vreps}) are non-negative.
This means that for the first factor in (\ref{deta0})
we take $j=0,1/2,\ldots$ and
for the second factor $j=1,3/2,\ldots$.
By taking account of this, we obtain
\begin{equation}
\Det'{\cal D}_{\cal A}
=(-ir\alpha(\sigma_0))\prod_{j=0}^\infty\prod_{|m|\leq j}(2j+2+2imu+ir\alpha(\sigma_0))^{2j+3}(2j-2imu-ir\alpha(\sigma_0))^{2j-1}.
\label{deta}
\end{equation}
The factor $-ir\alpha(\sigma_0)$ is inserted to remove the
unwanted contribution of the second factor with $j=0$.

\subsection{Fermions in vector multiplets}
The action for the fermion field $\lambda$ at the saddle point (\ref{saddle}) is
\begin{equation}
{\cal L}_\lambda=\frac{1}{r}\ol\lambda{\cal D}_\lambda\lambda,
\end{equation}
where the differential operator ${\cal D}_\lambda$ is given by
\begin{align}
{\cal D}_\lambda
&=
-r\gamma^m D_m
+\frac{i}{2v}
-\frac{1}{v}(1+iu\gamma_{\wh 3})r\alpha(\sigma_0)
\nonumber\\
&=-2\gamma_{\wh 1}L_1-2\gamma_{\wh 2}L_2-2v\gamma_{\wh 3}L_3
-iv
-\frac{1}{v}(1+iu\gamma_{\wh 3})r\alpha(\sigma_0)
\nonumber\\
&=\left(\begin{array}{cc}
-2v L_3-iv-\frac{1+iu}{v}r\alpha(\sigma_0) & -2L_{1-i2} \\
-2L_{1+i2} & 2vL_3-iv-\frac{1-iu}{v}r\alpha(\sigma_0)
\end{array}\right).
\end{align}
In the subspace with the quantum numbers
(\ref{restricted}),
this becomes $2\times 2$ matrix with the determinant
\begin{equation}
\det{\cal D}_\lambda
=(2j+1+ir\alpha(\sigma_0)+2imu)(2j+1-ir\alpha(\sigma_0)-2imu).
\end{equation}
The first and the second factor correspond
to the two irreducible representations in
\begin{equation}
(j,j)\otimes(0,\frac{1}{2})=(j,j-\frac{1}{2})\oplus(j,j+\frac{1}{2}).
\end{equation}
By taking the product over all possible quantum numbers
and ignoring a constant factor,
we obtain
\begin{equation}
\Det{\cal D}_\lambda
=\prod_j
\prod_{|m|\leq j-1/2}
(2j+1+ir\alpha(\sigma_0)+2imu)^{2j+1}
\prod_{|m|\leq j+1/2}
(2j+1-ir\alpha(\sigma_0)-2imu)^{2j+1}.
\end{equation}
Let us shift $j$ by $\pm 1/2$ so that the $SU(2)_R$ spin
of the two representations become the same
\begin{equation}
(j+\frac{1}{2},j)\oplus(j-\frac{1}{2},j).
\label{kambdasp}
\end{equation}
After the shift, the determinant becomes
\begin{equation}
\Det{\cal D}_\lambda=
\prod_{j=0}^\infty\prod_{|m|\leq j}
(2j+2+ir\alpha(\sigma_0)+2imu)^{2j+2}(2j-ir\alpha(\sigma_0)-2imu)^{2j}.
\label{detlambda}
\end{equation}

Combining (\ref{deta}), (\ref{detlambda}), and the Vandermonde determinant (\ref{vdm}), we obtain
\begin{align}
Z_{\rm vector}^{\rm 1-loop}(\sigma_0)
&=
\prod_{\alpha\in\Delta}\frac{\Det{\cal D}_\lambda}{\Det'{\cal D}_{\cal A}}\prod_{\alpha\in\Delta}\alpha(\sigma_0)
\nonumber\\
&=
\prod_{\alpha\in\Delta}\prod_j\prod_{|m|\leq j}
\frac{2j-ir\alpha(\sigma_0)-2imu}{2j+2+ir\alpha(\sigma_0)+2imu}.
\end{align}
If we set
\begin{equation}
j=\frac{p+q}{2},\quad
m=\frac{p-q}{2},
\label{mnpq}
\end{equation}
we obtain
\begin{align}
Z_{\rm vector}^{\rm 1-loop}(\sigma_0)
&=
\prod_{\alpha\in\Delta}
\prod_{p,q=0}^\infty
\frac{(1-iu)p+(1+iu)q+1-i(r\alpha(\sigma_0)-i)}{(1+iu)p+(1-iu)q+1+i(r\alpha(\sigma_0)-i)}
\nonumber\\
&=\prod_{\alpha\in\Delta}
s_b\left(\frac{r\alpha(\sigma_0)-i}{v}\right).
\end{align}
This is the same as the numerator in (\ref{zvc})
with $b$ and $x_0$ in (\ref{bxx0}).

\subsection{Bosons in chiral multiplets}
We can reduce the path integral with respect to chiral multiplets
to Gaussian integrals by sending the coefficient of ${\cal L}_{\rm chiral}$
to infinity.

Let us compute the contribution of bosonic fields in a chiral multiplet
with Weyl weight $\Delta$ belonging to a gauge representation $\cal R$.
The path integral of the auxiliary field $F$ gives constant, and we can neglect it.

Let us assume that $\phi$ is eigenmode of $\sigma_0$
and $\sigma_0\phi=\rho(\sigma_0)\phi$,
where $\rho$ is a weight in the representation $\cal R$.
At the saddle point (\ref{saddle}),
the scalar Lagrangian is
\begin{equation}
{\cal L}_\phi=\frac{1}{r^2}\phi^\dagger {\cal D}_\phi \phi,
\end{equation}
where the differential operator ${\cal D}_\phi$
is given by
\begin{equation}
{\cal D}_\phi=
-r^2 D_m D^m
-(ir\rho(\sigma_0)-\Delta+2)(ir\rho(\sigma_0)-\Delta)
-\frac{2u}{v}(ir\rho(\sigma_0)-\Delta+1) rD_{\wh 3}.
\end{equation}
We expand the scalar field
with ${\bf S}^3$ spherical harmonics $|j,m',m\rangle$.
These harmonics are eigenfunctions of
the Laplacian $D_m D^m$ and $D_{\wh 3}$.
\begin{align}
r^2 D_m D^m
|j,m',m\rangle
&=(-4j(j+1)-4u^2m^2)
|j,m',m\rangle
,\nonumber\\
rD_{\wh 3}
|j,m',m\rangle
&=2ivm
|j,m',m\rangle.
\end{align}
The eigenvalue of the differential operator ${\cal D}_\phi$
in the subspace defined by (\ref{restricted})
is
\begin{equation}
{\cal D}_\phi
=(2j+ir\rho(\sigma_0)-\Delta+2+2ium)(2j-ir\rho(\sigma_0)+\Delta-2ium).
\end{equation}
By taking the product over all possible quantum numbers,
we obtain the determinant of the differential operator
\begin{equation}
\Det{\cal D}_\phi=\prod_{j=0}^\infty\prod_{|m|\leq j}(2j+ir\rho(\sigma_0)-\Delta+2+2ium)^{2j+1}(2j-ir\rho(\sigma_0)+\Delta-2ium)^{2j+1}.
\label{detphi}
\end{equation}

\subsection{Fermions in chiral multiplets}
The linearized action of fermion fields $\psi$ and $\ol\psi$ at the
saddle point (\ref{saddle}) is
\begin{equation}
{\cal L}_\psi=\frac{1}{r}(\ol\psi{\cal D}_\psi\psi),
\end{equation}
where the differential operator ${\cal D}_\psi$ is
given by
\begin{equation}
{\cal D}_\psi
=-r\gamma^m D_m
+\frac{i}{2v}
-\frac{i(\Delta-ir\rho(\sigma_0))}{v}(1+iu\gamma^{\wh 3}).
\end{equation}
By using (\ref{diffgen}), we can rewrite this operator
in the algebraic form
\begin{align}
{\cal D}_\psi
&=
-\gamma_{\wh1+ i\wh2}L_{1- i2}
-\gamma_{\wh1- i\wh2}L_{1+ i2}
-2v\gamma_{\wh3}L_{3}
-iv
-\frac{i(\Delta-ir\sigma_0)}{v}(1+iu\gamma_{\wh 3})
\nonumber\\
&=
\left(\begin{array}{cc}
-2vL_3-iv
-\frac{i(\Delta-ir\rho(\sigma_0))}{v}(1+iu)
 & -2L_{1-i2} \\
-2L_{1+i2} & 2vL_3-iv
-\frac{i(\Delta-ir\rho(\sigma_0))}{v}(1-iu)
\end{array}\right).
\end{align}
In the vector space
with quantum numbers (\ref{restricted}),
this becomes $2\times 2$ matrix with the determinant
\begin{equation}
\det {\cal D}_\psi
=(2j+1+\Delta-ir\rho(\sigma_0)-2imu)(2j+1-\Delta+ir\rho(\sigma_0)+2imu).
\label{psidet}
\end{equation}
The two factors correspond to the two representations
in the irreducible decomposition
\begin{equation}
(j,j)\otimes(0,\frac{1}{2})=(j,j+\frac{1}{2})\oplus(j,j-\frac{1}{2}).
\label{irrdecomp}
\end{equation}
The first and the second factor in 
(\ref{psidet}) correspond to the first and the second
irreducible representations in (\ref{irrdecomp}).
By taking the product over all possible quantum numbers,
we obtain
\begin{equation}
\Det{\cal D}_\psi=
\prod_{j=0}^\infty\prod_{|m|\leq j}
(2j+\Delta-ir\rho(\sigma_0)-2ium)^{2j}
(2j-\Delta+ir\rho(\sigma_0)+2+2ium)^{2j+2},
\label{detpsi}
\end{equation}
where we shifted the quantum number $j$ so that
$SU(2)_R$ spins become $j$.

Combining (\ref{detphi}) and (\ref{detpsi}) we obtain
\begin{align}
Z_{\rm chiral}^{\rm 1-loop}
&=
\prod_{\rho\in{\cal R}}
\frac{\Det{\cal D}_\psi}{\Det{\cal D}_\phi}
\nonumber\\
&=
\prod_{\rho\in{\cal R}}
\prod_{j=0,1/2,\ldots}
\prod_{|m|\leq j}
\frac{2j-\Delta+2+ir\rho(\sigma_0)+2ium}
{2j+\Delta-ir\rho(\sigma_0)-2ium}.
\end{align}
After the variable change (\ref{mnpq})
we obtain
\begin{align}
Z_{\rm chiral}^{\rm 1-loop}
&=
\prod_{\rho\in{\cal R}}
\prod_{p,q=0}^\infty
\frac{(1+iu)p+(1-iu)q+1+i(r\rho(\sigma_0)+i\Delta-i)}
{(1-iu)p+(1+iu)q+1-i(r\rho(\sigma_0)+i\Delta-i)}
\nonumber\\
&=
1\Big/
\prod_{\rho\in{\cal R}}s_b\left(\frac{r\rho(\sigma_0)-i(1-\Delta)}{v}\right).
\end{align}
This is the contribution of one chiral multiplet
belonging to $\cal R$
with Weyl weight $\Delta$.
By multiplying the contributions of all chiral multiplets
we obtain the denominator in (\ref{zvc})
with $b$ and $x_{\Delta_I}$ in (\ref{bxx0}).

\section{4d to 3d}\label{4d3d.sec}
\subsection{4d theory}
As we mentioned in Introduction, the 3d theory we investigated
can be derived from a 4d theory by dimensional reduction.
In this section, we summarize the derivation of the
action and the transformation laws.

We first summarize the 4d conventions and notation.
We use Greek characters $\kappa,\lambda,\mu,\nu,\ldots,=1,2,3,4$ for 4d tangent indices,
and hatted ones $\wh\kappa,\wh\lambda,\wh\mu,\wh\nu,\ldots,=\wh1,\wh2,\wh3,\wh4$ for 4d local indices.
We use the Dirac's matrices
\begin{equation}
\gamma_{\wh m}=\left(\begin{array}{cc} 0 & \sigma_m \\
\sigma_m & 0 \end{array}\right),\quad
\gamma_{\wh 4}=\left(\begin{array}{cc} 0 & -i \\
i & 0 \end{array}\right).
\end{equation}
We call upper half of a Dirac spinor left components
and lower half right components.
We use unbarred and barred spinors for left-handed and right-handed spinors.

We start from a 4d theory defined in the background ${\bf S}^3\times {\bf R}$,
where ${\bf S}^3$ is a round sphere with radius $r$.
We use $g_{(0)}\in SU(2)$ and $x^4\in{\bf R}$ to parametrize ${\bf S}^3$
and ${\bf R}$, respectively.
The metric is
\begin{equation}
ds^2=r^2\left[
(\mu_{(0)}^1)^2
+(\mu_{(0)}^2)^2
+(\mu_{(0)}^3)^2
\right]+(dx^4)^2,
\label{cylmet}
\end{equation}
where $\mu_{(0)}^a$ is the left-invariant $1$-form
defined by
\begin{equation}
2\mu_{(0)}^aT_a=g_{(0)}^{-1}dg_{(0)}.
\end{equation}
For later convenience, we define vector fields $h$ and $t_a$ ($a=1,2,3$) by
\begin{equation}
h=\left(\frac{\partial}{\partial x^4}\right)_{g_{(0)}},\quad
t_a g_{(0)}=2g_{(0)}T_a.
\label{nt}
\end{equation}
$h$ is the translation along ${\bf R}$, and
$t_a$ are the dual basis to $\mu^a$.
By definition $(t_a,\mu^b)=\delta_a^b$.

This manifold admits four left-handed Killing spinors $\epsilon_i$
and four right-handed Killing spinors $\ol\epsilon_i$ ($i=1,2,3,4$).
They have the quantum numbers shown in Table
\ref{table:fourspinors},
\begin{table}[htb]
\caption{Quantum numbers of eight Killing spinors
in ${\bf S}^3\times{\bf R}$}
\label{table:fourspinors}
\begin{center}
\begin{tabular}{ccccccccc}
\hline
\hline
& $\epsilon_1$ & $\epsilon_2$ & $\epsilon_3$ & $\epsilon_4$
& $\ol\epsilon_1$ & $\ol\epsilon_2$ & $\ol\epsilon_3$ & $\ol\epsilon_4$ \\
\hline
$R$
& $1$ & $1$ & $1$ & $1$
& $-1$ & $-1$ & $-1$ & $-1$ \\
$T_3^L$
& $-\frac{i}{2}$ & $\frac{i}{2}$ & $0$ & $0$
& $\frac{i}{2}$ & $-\frac{i}{2}$ & $0$ & $0$ \\
$T_3^R$
& $0$ & $0$ & $-\frac{i}{2}$ & $\frac{i}{2}$
& $0$ & $0$ & $\frac{i}{2}$ & $-\frac{i}{2}$ \\
$D=-r\partial_4$
 & $\frac{1}{2}$ & $\frac{1}{2}$ & $-\frac{1}{2}$ & $-\frac{1}{2}$
& $-\frac{1}{2}$ & $-\frac{1}{2}$ & $\frac{1}{2}$ & $\frac{1}{2}$ \\
\hline
\end{tabular}
\end{center}
\end{table}
and satisfy the Killing equations
\begin{align}
&D_\mu\epsilon_{1/2}=-\frac{1}{2r}\gamma_\mu\sla h\epsilon_{1/2},\quad
D_\mu\epsilon_{3/4}=+\frac{1}{2r}\gamma_\mu\sla h\epsilon_{3/4},
\nonumber\\
&D_\mu\ol\epsilon_{1/2}=+\frac{1}{2r}\gamma_\mu\sla h\ol\epsilon_{1/2},\quad
D_\mu\ol\epsilon_{3/4}=-\frac{1}{2r}\gamma_\mu\sla h\ol\epsilon_{3/4}.
\label{4dkilling}
\end{align}
where $\sla h=h^\mu\gamma_\mu$.

Because the background (\ref{cylmet}) is conformally flat,
we can easily obtain the supersymmetry transformation laws
from those in the flat spacetime
by Weyl transformation.
The transformation laws for vector multiplets
are
\begin{align}
\delta A_\mu&=
i(\epsilon\gamma_\mu\ol\lambda)
-i(\ol\epsilon\gamma_\mu\lambda)
,\nonumber\\
\delta\lambda&=
\frac{i}{2}\gamma^{\mu\nu}\epsilon F_{\mu\nu}+D\epsilon,\nonumber\\
\delta\ol\lambda&=
-\frac{i}{2}\gamma^{\mu\nu}\ol\epsilon F_{\mu\nu}+D\ol\epsilon,\nonumber\\
\delta D&=
-(\epsilon\gamma^\mu D_\mu\ol\lambda)-(\ol\epsilon\gamma^\mu D_\mu\lambda).
\label{4dvectortr}
\end{align}
Transformation laws for chiral multiplets are
\begin{align}
\delta \phi&=\sqrt2(\epsilon\psi),\nonumber\\
\delta \phi^\dagger&=\sqrt2(\ol\epsilon\ol\psi),\nonumber\\
\delta\psi&=
-\sqrt{2}\gamma^\mu\ol\epsilon D_\mu\phi
+\sqrt{2}\epsilon F
-\frac{\Delta}{\sqrt2}\gamma^\mu D_\mu\ol\epsilon\phi
,\nonumber\\
\delta\ol\psi&=
-\sqrt{2}\gamma^\mu\epsilon D_\mu\phi^\dagger
+\sqrt{2}\ol\epsilon F^\dagger
-\frac{\Delta}{\sqrt2}\gamma^\mu D_\mu\epsilon\phi^\dagger
,\nonumber\\
\delta F&=
-\sqrt{2}(\ol\epsilon\gamma^\mu D_\mu\psi)
-2(\ol\epsilon\ol\lambda)\phi
-\frac{\Delta-1}{\sqrt2}D_\mu\ol\epsilon\gamma^\mu\psi,\nonumber\\
\delta F^\dagger&=
-\sqrt{2}(\epsilon\gamma^\mu D_\mu\ol\psi)
-2\phi^\dagger(\epsilon\lambda)
-\frac{\Delta-1}{\sqrt2}D_\mu\epsilon\gamma^\mu\ol\psi.
\label{4dcfchiraltr}
\end{align}

The kinetic Lagrangians for vector and chiral multiplets can be obtained
in the same way as in 3d
\begin{equation}
(\ol\epsilon_1\ol\epsilon_2){\cal L}_{\rm YM}^{(4d)}=
-\frac{1}{4}\delta(\ol\epsilon_1)
\delta(\ol\epsilon_2)\tr(\ol\lambda\ol\lambda),\quad
(\ol\epsilon_1\ol\epsilon_2){\cal L}_{\rm chiral}^{(4d)}=
-\frac{1}{2}\delta(\ol\epsilon_1)
\delta(\ol\epsilon_2)(\phi^\dagger F).
\end{equation}
The explicit form of these kinetic Lagrangians is
\begin{align}
{\cal L}_{\rm YM}^{(4d)}=
{\cal L}_{\cal A}^{(4d)}
+{\cal L}_\lambda^{(4d)}
-\frac{1}{2}\tr D^2,\quad
{\cal L}_{\rm chiral}^{(4d)}=
{\cal L}_\phi^{(4d)}
+{\cal L}_\psi^{(4d)}
-F^\dagger F,
\end{align}
where
\begin{align}
{\cal L}^{(4d)}_{\cal A}
&=\tr\frac{1}{2}{\cal F}_{\wh m}^{(-)}{\cal F}_{\wh m}^{(-)}
,\nonumber\\
{\cal L}^{(4d)}_\lambda
&=
-\tr(\ol\lambda\gamma^\mu D_\mu\lambda)
,\nonumber\\
{\cal L}_\phi^{(4d)}
&=
   -\phi^\dagger D_\mu D^\mu \phi
   +\phi^\dagger D\phi
   -\frac{\Delta^2-2\Delta}{r^2}\phi^\dagger\phi
   -\frac{2(\Delta-1)}{r}h^\mu\phi^\dagger D_\mu\phi
,\nonumber\\
{\cal L}_\psi^{(4d)}
&=
   -(\ol\psi\gamma^\mu D_\mu\psi)
   -\frac{\Delta-1}{r}h^\mu(\ol\psi\gamma_\mu\psi)
   -\sqrt2\phi^\dagger(\lambda\psi)
   -\sqrt2(\ol\psi\ol\lambda)\phi.
\end{align}
${\cal F}_{\wh m}^{(\pm)}$ are defined by
\begin{equation}
{\cal F}_{\wh m}^{(\pm)}=\frac{1}{2}\epsilon_{\wh m\wh p\wh q}F_{\wh p\wh q}\pm F_{\wh m\wh 4}.
\label{vin4d}
\end{equation}

\subsection{Killing spinors and twisted compactification}
To obtain 3d theory, we need to compactify the ${\bf R}$ direction.
This is realized by imposing the condition
\begin{equation}
{\cal O}\Phi=\Phi,
\label{ophiphi}
\end{equation}
on all fields $\Phi$ in the theory,
where ${\cal O}$ is an operator containing shift along $x^4$ and
additional twists.
To keep some of supersymmetries unbroken,
we should choose ${\cal O}$ which keep the corresponding Killing spinors
invariant.
Our choice is
\begin{equation}
{\cal O}=q^{D-\frac{1}{2}R_0-2uT^R_3},\quad
q=e^{-\beta},
\label{ouro}
\end{equation}
where $D=-r\partial_4$ is the $x^4$-translation,
and
$\beta$ is the period of the ${\bf S}^1$ compactification
divided by the ${\bf S}^3$ radius $r$.
$R_0$ is an R-symmetry.
This is not the R-symmetry in the superconformal algebra,
but one that does not rotate the dynamical scalar components of chiral multiplets.
\begin{equation}
R_0(\phi)=R_0(\phi^\dagger)=0,\quad
R_0(\ol\epsilon)=R_0(\psi)=-1,\quad
R_0(\lambda)=+1.
\end{equation}
This twist preserves four supersymmetries out of eight corresponding
to $\epsilon_1$, $\epsilon_2$, $\ol\epsilon_1$, and $\ol\epsilon_2$.
Note that when $u\neq 0$ this compactification breaks $SU(2)_R$ to $U(1)_r$.

\begin{figure}[htb]
\centerline{\includegraphics{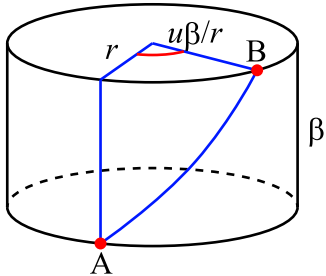}}
\caption{Twisted compactification of ${\bf S}^3\times{\bf R}$ is shown.
Points A and B are identified.}
\label{cyl.eps}
\end{figure}

The constraint (\ref{ophiphi})
with the operator ${\cal O}$ in (\ref{ouro})
implies the following identification of the points
\begin{equation}
(g_{(0)} e^{\frac{2u}{r}\beta T_3^R},x^4+\beta)
\sim (g_{(0)},x^4).
\label{identification}
\end{equation}
(Figure \ref{cyl.eps}.)
We take the small radius limit $\beta\rightarrow 0$,
and get rid of all Kaluza-Klein modes except
the lowest one for each field to obtain 3d theory.
This reduction is realized by imposing the constraint
\begin{equation}
\left(D-2uT_3^R-\frac{1}{2}R_0\right)\Phi=0
\label{opconst}
\end{equation}
on all fields.
By using the vector fields in (\ref{nt}), we can rewrite
this as the differential equation
\begin{equation}
\left(-{\cal L}_{rh+ut_3}-\frac{1}{2}R_0\right)
\Phi=0.
\label{psiconsya}
\end{equation}
The constraint (\ref{psiconsya})
determines the $x^4$ dependence of fields
from their values on the $x^4=0$ slice.

It is convenient to
perform
the coordinate transformation
\begin{equation}
(g,x^4)=(g_{(0)}e^{-\varphi(x^4)T_3^R},x^4),\quad
\varphi(x^4)=\frac{2u}{r}x^4.
\end{equation}
In the new coordinate system,
the identification (\ref{identification}) is simplified as
\begin{equation}
(g,x^4+\beta)
\sim (g,x^4).
\end{equation}
The metric in the new coordinate system is
\begin{align}
ds^2
&=
r^2\left[
(\mu^1)^2+
(\mu^2)^2
+\frac{1}{v^2}(\mu^3)^2\right]
+v^2(dx^4+V)^2,
\end{align}
where
$\mu^a$ are defined in (\ref{lioneform}), and the 1-form $V$ is
\begin{equation}
V=\frac{ru}{v^2}\mu^3.
\end{equation}
After dimensional reduction, we obtain
squashed sphere (\ref{sqmetric}) with the background
graviphoton field $V$.
Note that this $V$ is the same as (\ref{graviphoton}).

We use the 4d vielbein
\begin{equation}
E^{\wh 1}=r\mu^1,\quad
E^{\wh 2}=r\mu^2,\quad
E^{\wh 3}=\frac{r}{v}\mu^3,\quad
E^{\wh 4}=v(dx^4+V).
\label{twistframe}
\end{equation}
In general, the components of the 4d spin connection
$\Omega$ of the 4d manifold with the metric
\begin{equation}
ds^2=E^{\wh \mu}E^{\wh \mu}=e^{\wh m}e^{\wh m}+v^2(dx^4+V)^2,
\label{s1compact}
\end{equation}
are
\begin{equation}
\Omega_{\wh m-\wh n\wh p}=\omega_{\wh m-\wh n\wh p},\quad
\Omega_{\wh 4-\wh m\wh n}=-\frac{v}{2}(dV)_{\wh m\wh n},\quad
\Omega_{\wh m-\wh 4\wh n}=-\frac{v}{2}(dV)_{\wh m\wh n},\quad
\Omega_{\wh 4-\wh 4\wh m}=0,
\end{equation}
where
$\omega$ is the spin connection of the 3d manifold with the metric $ds^2=e^{\wh m}e^{\wh m}$.
By using the explicit form of the graviphoton field $V$,
we obtain
\begin{equation}
\Omega_{\wh 4-\wh 1\wh 2}
=\Omega_{\wh 1-\wh 4\wh 2}
=-\Omega_{\wh 2-\wh 4\wh 1}
=-\frac{u}{vr}.
\label{spinc}
\end{equation}

The components of the vielbein are
\begin{equation}
\left(\begin{array}{cc}
E_m{}^{\wh n} & E_m{}^{\wh 4} \\
E_4{}^{\wh n} & E_4{}^{\wh 4}
\end{array}\right)
=
\left(\begin{array}{cc}
e_m{}^{\wh n} & ue_m^{\wh 3} \\
0 & v
\end{array}\right),\quad
\left(\begin{array}{cc}
E_{\wh m}{}^n & E_{\wh m}{}^4 \\
E_{\wh 4}{}^n & E_{\wh 4}{}^4
\end{array}\right)
=
\left(\begin{array}{cc}
e_{\wh m}{}^n & -\frac{u}{v}\delta_{\wh m}^{\wh 3} \\
0 & \frac{1}{v}
\end{array}\right).
\label{eandE}
\end{equation}

In the new coordinate system, the vector field appearing in the constraint
(\ref{psiconsya}) is
\begin{equation}
rh+ut_3=r\frac{\partial}{\partial x^4}.
\end{equation}
By using this and the spin connection in (\ref{spinc}),
the constraint (\ref{psiconsya}) is simplified as
\begin{equation}
\left(-r\frac{\partial}{\partial x^4}-\frac{R_0}{2}\right)\Phi=0.
\label{constdigg}
\end{equation}

\subsection{Dimensional reduction}
We define 3d fields as the restriction of the corresponding 4d fields
on the slice $x^4=0$.
For a 4d left-handed (right-handed) spinor field,
we take two components of the left-handed (right-handed) part
of the 4d field as the 3d field.
For example, for the left-handed spinor field $\lambda^{(4d)}$
we define the corresponding 3d field by
\begin{equation}
\lambda^{(4d)}|_{x^4=0}=\left(\begin{array}{c} \lambda^{(3d)} \\ 0 \end{array}\right).
\end{equation}
A 4d gauge field $A^{(4d)}=A_\mu^{(4d)}dx^\mu$ is decomposed into 3d gauge field $A^{(3d)}=A_m^{(3d)}dx^m$ and 3d
adjoint scalar field $\sigma$ by
\begin{equation}
A_\mu^{(4d)}|_{x^4=0}dx^\mu=A^{(3d)}+\sigma dx^4.
\end{equation}

To obtain 3d Lagrangians and transformation laws,
we need to rewrite the 4d covariant derivatives in terms of 3d ones.
By using the explicit form of the vielbein and spin connection,
we obtain
\begin{align}
\frac{1}{2}E_{\wh m}^\mu\Omega_{\mu-\wh\kappa\wh \lambda}S_{\kappa\lambda}
&=\frac{1}{2}e_{\wh m}^n\omega_{n-\wh k\wh l}S_{kl}
-\frac{u}{vr}\epsilon_{\wh m\wh n\wh 3}S_{4n},
\nonumber\\
\frac{1}{2}E_{\wh 4}^\mu\Omega_{\mu-\wh \kappa\wh \lambda}S_{\kappa\lambda}
&=-\frac{u}{vr}S_{12},
\end{align}
where $S_{\mu\nu}$ are 4d spin operators.
With these relations and the constraint (\ref{constdigg}),
we can easily obtain
\begin{align}
D_{\wh m}^{(4d)}
&=D_{\wh m}^{(3d)}
-\frac{u}{vr}\epsilon_{\wh m\wh 3\wh n}S_{n4}
+\frac{u}{vr}\delta_{\wh m\wh 3}\left(\frac{R_0}{2}+ir\sigma\right),\nonumber\\
D_{\wh 4}^{(4d)}
&=-\frac{1}{vr}\left(\frac{R_0}{2}+ir\sigma\right)-\frac{u}{vr}S_{12}.
\label{d4dd3d}
\end{align}
By using these relations, it is straightforward to obtain
3d supersymmetry transformation laws and 3d Lagrangians from
4d ones.
(The Chern-Simons term cannot be obtained from 4d Lagrangian,
and we need to construct it by, for example, Noether procedure.)
We will not explain them in detail.
We only demonstrate the derivation of the 3d Killing equations (\ref{3dkilling}).
Our compactification preserves the Killing spinors $\epsilon_1$, $\epsilon_2$, $\ol\epsilon_1$, and $\ol\epsilon_2$.
They satisfy the 4d Killing equations
\begin{equation}
D_{\wh\mu} \epsilon=-\frac{1}{2r}\gamma_{\wh\mu}\sla h\epsilon,\quad
D_{\wh\mu} \ol\epsilon=\frac{1}{2r}\gamma_{\wh\mu}\sla h\ol\epsilon.
\label{4dkil}
\end{equation}
For $\wh\mu=\wh m$, the left hand side of these equations are rewritten by (\ref{d4dd3d}) as
\begin{align}
D_{\wh m}^{(4d)}\epsilon
&=D_{\wh m}^{(3d)}\epsilon
-\frac{u}{2vr}\epsilon_{\wh m\wh 3\wh n}\gamma_{\wh n\wh 4}\epsilon
+\frac{u}{2vr}\delta_{\wh m\wh 3}\epsilon
=D_{\wh m}^{(3d)}\epsilon
+\frac{u}{2vr}\gamma_{\wh 3}\gamma_{\wh m}\epsilon,\nonumber\\
D_{\wh m}^{(4d)}\ol\epsilon
&=D_{\wh m}^{(3d)}\ol\epsilon
-\frac{u}{2vr}\epsilon_{\wh m\wh 3\wh n}\gamma_{\wh n\wh 4}\ol\epsilon
-\frac{u}{2vr}\delta_{\wh m\wh 3}\ol\epsilon
=D_{\wh m}^{(3d)}\ol\epsilon
-\frac{u}{2vr}\gamma_{\wh 3}\gamma_{\wh m}\ol\epsilon.
\label{d4de}
\end{align}
The right hand side of the equations in (\ref{4dkil}) are rewritten as
\begin{align}
-\frac{1}{2r}\gamma_{\wh m}\sla h\epsilon
&=
-\frac{1}{2vr}\gamma_{\wh m}(\gamma_{\wh 4}-u\gamma_{\wh 3})\epsilon
=
-\frac{i}{2vr}\gamma_{\wh m}\epsilon
+\frac{u}{2vr}\gamma_{\wh m}\gamma_{\wh 3}\epsilon
,\nonumber\\
\frac{1}{2r}\gamma_{\wh m}\sla h\ol\epsilon
&=
\frac{1}{2vr}\gamma_{\wh m}(\gamma_{\wh 4}-u\gamma_{\wh 3})\ol\epsilon
=
-\frac{i}{2vr}\gamma_{\wh m}\ol\epsilon
-\frac{u}{2vr}\gamma_{\wh m}\gamma_{\wh 3}\ol\epsilon.
\label{rhhkeq}
\end{align}
where we used the fact that the vector field $h$ has the components
\begin{equation}
h^{\wh\mu}=\left(0,0,-\frac{u}{v},\frac{1}{v}\right).
\end{equation}
Combining (\ref{d4de}) and 
(\ref{rhhkeq}), we obtain the 3d Killing equations (\ref{3dkilling}).

\section{Large $N$ limit}\label{largen.sec}
In this section we investigate the free energy
$F = -\log Z$ of large $N$ gauge theories which are expected to have
M-theory dual.
We consider a quiver gauge theory with gauge group
\begin{equation}
G=\prod_{a=1}^{n_G} U(N)_a.
\end{equation}
In this case the traces in (\ref{csandfi}) are expressed as
linear combinations of the traces for $U(N)_a$ gauge groups,
\begin{equation}
\tr_{\rm CS}
=\sum_{a=1}^{n_G}\frac{k_a}{2\pi}\tr_a,\quad
\tr_{\rm FI}
=\sum_{a=1}^{n_G}\frac{\zeta_a}{vr}\tr_a,
\label{kaandzetaa}
\end{equation}
where $\tr_a$ is the trace over the $U(N)_a$ fundamental representation.
The coefficients $k_a$ and $\zeta_a$ are Chern-Simons levels and
FI parameters, respectively.
The Chern-Simons parameters $k_a$ must be integers.
The normalization of the FI parameters $\zeta_a$ is chosen for later convenience.

It is pointed out in \cite{Herzog:2010hf} that in order to
obtain the leading term of the free energy in the $1/N$ expansion,
we do not have to perform the integral over $\sigma_0$ in (\ref{zintf}).
We only need to determine the minimum value of the integrand
of (\ref{zintf}).
Namely, we obtain the free energy  by minimizing
\begin{equation}
F(\sigma_0)=S^{\rm cl}(\sigma_0)-\log Z^{\rm 1-loop}(\sigma_0).
\end{equation}
It is convenient to decompose this into three parts:
the classical action $F_1=S^{\rm cl}$,
the $1$-loop contribution of vector and bi-fundamental chiral multiplets $F_2$,
and the $1$-loop contribution of fundamental and anti-fundamental chiral multiplets $F_3$.

From (\ref{csficl}), the classical action $F_1$ is
\begin{equation}
F_1=S_{\rm CS}^{\rm cl}+S_{\rm FI}^{\rm cl}
=\sum_{a=1}^{n_G}\sum_{i=1}^N\left(
\frac{\pi i}{v^2}k_a \lambda_{a,i}^2
+\frac{4\pi^2 i}{v^2}\zeta_a\lambda_{a,i}\right),
\label{fdisc1}
\end{equation}
where $\lambda_{a,i}$ are diagonal components of the expectation value of
the $U(N)_a$ adjoint scalar field rescaled by $r$
\begin{equation}
r(\sigma_0)_a= \diag\{\lambda_{a,j}\}.
\end{equation}

$F_2$ is the $1$-loop contribution of
vector multiplets and bi-fundamental chiral multiplets.
It is given by
\begin{align}
F_2=& -\sum_{a=1}^{n_G} \sum_{j\neq k}
f_b\left(\frac{1}{v}\left(\lambda_{a,j}-\lambda_{a,k}-i\right)\right)
\nonumber\\
&+
\sum_{I\in{\rm bi-fund}} \sum_{j,k}
f_b\left(\frac{1}{v}\left(\lambda_{h(I),j}-\lambda_{t(I),k}-i\left(1-\Delta_I\right)\right)\right),
\label{fdisc2}
\end{align}
where $f_b(z)=\log s_b(z)$ and $b$ is the parameter related to the squashing parameter by (\ref{bxx0}).
The first line and the second line are contribution of vector and bi-fundamental chiral multiplets, respectively.
$I$ runs over all bi-fundamental chiral multiplets.
We use $h(I)$ and $t(I)$ to represent
the $U(N)$ factors at the head and the tail of the arrow
corresponding to the chiral multiplet $I$
in the quiver diagram.
Namely, a chiral multiplet $I$ belongs to the bi-fundamental representation
$(N_{h(I)},\ol N_{t(I)})$.
Adjoint chiral multiplets are treated as bi-fundamental chiral
multiplets with $h(I)=t(I)$,
and their contribution is also included in $F_2$.

The contribution of fundamental and anti-fundamental chiral multiplets
is denoted by $F_3$, and given by
\begin{align}
F_3=&
\sum_{I\in{\rm fund}} \sum_{j}
f_b\left(\frac{1}{v}\left(\lambda_{h(I),j}-i\left(1-\Delta_I\right)\right)\right)
\nonumber\\
&+
\sum_{I\in{\rm anti-fund}} \sum_{j}
f_b\left(\frac{1}{v}\left(-\lambda_{h(I),j}-i\left(1-\Delta_I\right)\right)\right),
\label{fdisc3}
\end{align}
where $I\in$fund and $I\in$anti-fund mean that the
index $I$ runs over fundamental and anti-fundamental chiral multiplets, respectively.

In \cite{Herzog:2010hf} the minimum points are determined numerically in
some models,
and the eigenvalue distribution is found to
behave in the large $N$ limit as
\begin{equation}
\lambda_{a,j} = N^\alpha x_j+iy_{a,j} ,
\label{dist}
\end{equation}
where $x_j$ and $y_{a,j}$ are real numbers,
and $\alpha$ is a certain constant in the region $0<\alpha<1$.
Note that $x_j$ are common for all $U(N)_a$ factors.
In \cite{Jafferis:2011zi}, the analysis is extended to a large class of
quiver gauge theories, and it is shown that we can consistently
determine
the free energy
proportional to $N^{3/2}$
based on the ansatz (\ref{dist})
if the theory satisfies the following conditions.
\begin{itemize}
\item[(A)]
The theory is non-chiral.
This means that the number of bi-fundamental chiral multiplets
transforming in $(N,\ol N)$ of the gauge group $U(N)_a \times U(N)_b$
is the same as that in $(\ol N,N)$.

\item[(B)]

The Weyl weights of chiral multiplets satisfy
\begin{equation}
\sum_{I\in a}(1-\Delta_I)-2=0,\quad
\forall a,
\label{anomaly}
\end{equation}
where $\Delta_I$ is the Weyl weight of the bi-fundamental field $I$.
The sum is taken over all bi-fundamental fields coupled by $U(N)_a$.
A $U(N)_a$ adjoint chiral multiplet should be included twice.
Fundamental and anti-fundamental fields should not be included.

\item[(C)]

The total number of fundamental fields and anti-fundamental fields should be the same.
Note that this condition is not imposed for each $U(N)_a$ factor.
The numbers of fundamental and anti-fundamental fields for each $U(N)_a$ factor
may be different.
Only the total numbers matter.

\item[(D)]

Chern-Simons levels sum up to zero:
\begin{align}
 \sum_{a=1}^{n_G} k_a = 0 .
\end{align}

\end{itemize}

In \cite{Jafferis:2011zi}, it is shown that
the free energy of theories
satisfying these condition defined on round ${\bf S}^3$
is proportional to $N^{3/2}$.
We generalize it to theories in the squashed ${\bf S}^3$.
We follow the prescription proposed in \cite{Jafferis:2011zi}.

The first step to determine the free energy in the large $N$ limit
is to rewrite the summations in
(\ref{fdisc1}), (\ref{fdisc2}), and (\ref{fdisc3})
by integrals.
We define the density function $\rho(x)$ by
\begin{equation}
\rho(x)=\frac{1}{N}\sum_{j=1}^N \delta(x-x_j).
\end{equation}
By definition, $\rho$ satisfies the normalization condition
\begin{equation}
\int_{x_{\rm min}}^{x_{\rm max}} \rho(x)dx=1.
\label{rhonorm}
\end{equation}
In the large $N$ limit, we can treat $\rho$ as a continuous function of
$x$.
Similarly, we replace $y_{a,i}$ by functions $y_a(x)$.
The classical action contribution $F_1$ is rewritten in the continuous form
as
\begin{equation}
F_1
=N\sum_{a=1}^{n_G}\int_{x_{\rm min}}^{x_{\rm max}} dx \rho\left(
\frac{\pi i}{v^2}k_a \lambda_a^2
+\frac{4\pi^2 i}{v^2}\zeta_a\lambda_a\right).
\label{f1cont}
\end{equation}
We substitute the continuous form of 
(\ref{dist})
\begin{equation}
\lambda_a(x) = N^\alpha x+iy_a(x)
\label{lxy}
\end{equation}
into (\ref{f1cont}).
Thanks to the condition (D), $N^{1+2\alpha}$ terms cancel,
and the leading terms are proportional to $N^{1+\alpha}$.
If we ignore sub-leading terms, we obtain
\begin{equation}
F_1
=\frac{\pi N^{1+\alpha}}{v^2}
\sum_{a=1}^{n_G}\int_{x_{\rm min}}^{x_{\rm max}} dx \rho x
(-2 k_ay_a+4\pi i\zeta_a).
\label{f1contb}
\end{equation}

$F_2$ in (\ref{fdisc2}) is
rewritten as
\begin{align}
F_2=&
-
N^2 \int_{x_{\rm min}}^{x_{\rm max}}  dx\int_{x_{\rm min}}^{x_{\rm max}}  dx' \rho\rho'
\sum_a
f_b\left(\frac{1}{v}(\lambda_a-\lambda_a'-i)
\right)
\nonumber\\
&
+
N^2 \int_{x_{\rm min}}^{x_{\rm max}}dx \int_{x_{\rm min}}^{x_{\rm max}} dx' \rho\rho'
\sum_{I\in{\rm adj}}
f_b\left(\frac{1}{v}(\lambda_{h(I)}-\lambda'_{t(I)}-i(1-\Delta_I)\right),
\label{f2cont}
\end{align}
where $\rho'\equiv \rho(x')$ and $\lambda_a'\equiv\lambda_a(x')$.
The key idea to rewrite these double integrals to
tractable form is that
if $x\neq x'$ we can replace the function $f_b$
by its asymptotic form
\begin{equation}
f_b^{\rm asym}(z)=
i\pi\left(\frac{z^2}{2}+\frac{b^2+b^{-2}}{24}\right)\sign(x),
\label{fasym0}
\end{equation}
because the real part of eigenvalues $\lambda$ scales as $N^{\alpha}$
in the large $N$ limit.
We call this ``long range potential.''
The contribution from $x=x'$ should be taken separately
as a ``short range potential''
proportional to $\delta(x-x')$.
In the large $N$ limit, we can replace the function $f_b(z)$
by the sum of long range and short range potentials
\begin{equation}
f_b(x+iy)\rightarrow f_b^{\rm asym}(x+iy)+\delta(x)g_b(y),
\end{equation}
where the function $g_b(y)$ is given by
\begin{align}
g_b(y)
&=
\frac{\pi}{3}y^3-\frac{\pi}{12}(b^2+b^{-2})y.
\label{funcg0}
\end{align}
See Appendix \ref{asym.sec} for a derivation of
(\ref{fasym0}) and (\ref{funcg0}).

Let us consider the contribution of the long-range potential
in (\ref{f2cont}).
$f^{\rm asym}_b(z)$ is a quadratic function of $z$, and after substitution of
(\ref{lxy}),
(\ref{f2cont}) contains terms of order $N^{2+2\alpha}$, $N^{2+\alpha}$, and
$N^2$.
To obtain the free energy of order $N^{3/2}$, all these terms should cancel.
This is indeed the case.
We can easily show that the contribution of long range potential in $F_2$
cancel 
due to the conditions (A) and (B).
As a result, only the short range potential contributes to $F_2$.
Because the short range potential contains $\delta$-function,
we can perform one of integrals.
After the $x'$ integral, $F_2$ is given by
\begin{align}
F_2
&=vN^{2-\alpha} \int_{x_{\rm min}}^{x_{\rm max}} dx \rho^2 \left[ \sum_{I\in{\rm bi-fund}} g_b \left( \frac{1}{v} \left(y_I - \left( 1-\Delta_I \right) \right) \right)
-\sum_{a=1}^{n_G} g_b \left(-\frac{1}{v}\right)
\right],
\label{f2gb}
\end{align}
where we defined
\begin{equation}
y_I=y_{h(I)}-y_{t(I)}.
\end{equation}
By using (\ref{anomaly}), we can rewrite the second term
in the brackets
in (\ref{f2gb}) as the summation over bi-fundamental chiral multiplets
\begin{align}
F_2
&=vN^{2-\alpha} \int_{x_{\rm min}}^{x_{\rm max}} dx \rho^2  \sum_{I\in{\rm bi-fund}}\left[ g_b \left( \frac{1}{v} \left(y_I - \left( 1-\Delta_I \right) \right) \right)
-(1-\Delta_I) g_b \left(-\frac{1}{v}\right)
\right]
\nonumber\\
&=\frac{\pi N^{2-\alpha}}{v^2} \int_{x_{\rm min}}^{x_{\rm max}} dx \rho^2  \sum_{I\in{\rm bi-fund}}
\frac{1}{3}(y_I+\Delta_I)(y_I-1+\Delta_I)(y_I-2+\Delta_I)
.
\end{align}
To obtain the second line we used $\sum_I y_I=0$ following from the condition (A).

The continuous form of the contribution of fundamental and anti-fundamental fields,
(\ref{fdisc3}), is
\begin{align}
F_3 =&
N \int_{x_{\rm min}}^{x_{\rm max}} dx \rho
\sum_{I\in{\rm fund}}
f_b\left(\frac{\lambda_{h(I)}-i(1-\Delta_I)}{v}\right)
\nonumber\\
&+
N \int_{x_{\rm min}}^{x_{\rm max}} dx \rho
\sum_{I\in{\rm anti-fund}}
f_b\left(\frac{-\lambda_{t(I)}-i(1-\Delta_I)}{v}\right).
\end{align}
Order $N^{1+2\alpha}$ terms
in the long range potential contribution
cancel by the condition (C),
and the leading non-vanishing terms in $F_3$ are of order $N^{1+\alpha}$.
The contribution of the short range potential is
of order $N^{1-\alpha}$, and we can neglect them.
The leading terms in $F_3$ are
\begin{align}
F_3 =&
\frac{\pi N^{1+\alpha}}{v^2} \int_{x_{\rm min}}^{x_{\rm max}} dx \rho
\sum_{I\in{\rm fund}}
|x| (1-\Delta_I-y_{h(I)})
\nonumber\\
&+
\frac{\pi N^{1+\alpha}}{v^2} \int_{x_{\rm min}}^{x_{\rm max}} dx \rho
\sum_{I\in{\rm anti-fund}}
|x| (1-\Delta_I+y_{t(I)}).
\end{align}

Now we have succeeded in writing all the contributions
to the free energy as one-dimensional integral.
$F_1$ and $F_3$ are proportional to $N^{1+\alpha}$,
and $F_2$ is proportional to $N^{2-\alpha}$.
To obtain minimum point, these should balance,
and this require $\alpha=1/2$.
In this case, the free energy is proportional to $N^{3/2}$,
as is expected from the analysis on the gravity side of AdS/CFT.

Let us focus on the dependence on the squashing parameter $v$.
We find that in all terms of order $N^{3/2}$
the $v$ dependence is factored out as the factor $1/v^2$.
(For the contribution of FI terms, this is the case when we adopt the
normalization of FI parameters in (\ref{kaandzetaa}).)
Therefore, the free energy obtained by minimizing the
$x$-integral is always $1/v^2$ times as that for round ${\bf S}^3$:
\begin{equation}
F_{\rm squashed}=\frac{1}{v^2}F_{\rm round}.
\end{equation}
This fact guarantees that the R charge at the IR fixed point
obtained by extremizing $Z$ does not depend
on the squashing parameter.

\section{Conclusions}\label{conc.sec}
We investigated ${\cal N}=2$ supersymmetric theories on squashed sphere with $SU(2)_L\times U(1)_r$ isometry.
The theories have four supercharges,which are transformed by $SU(2)_L$ isometry
as a pair of doublets.
We constructed supersymmetry transformation laws and Lagrangians
by using ${\bf S}^1$ compactification of 4d theory.
Although the metric of the squashed sphere
is the same as that of the $SU(2)_L\times U(1)_r$ symmetric squashing in
\cite{Hama:2011ea}, the supersymmetry group is different.
We computed the partition function by using localization,
and showed that it depends on the
squashing parameter in a non-trivial way.

We also computed the free energy of large $N$ quiver gauge theories on
the squashed ${\bf S}^3$.
We considered a class of quiver gauge theories studied in \cite{Jafferis:2011zi},
whose partition function on round ${\bf S}^3$ scales as $N^{3/2}$.
We confirmed that the free energy on squashed ${\bf S}^3$
is proportional to $N^{3/2}$ as well,
and
the $v$ dependence is factored out as the additional factor $1/v^2$
regardless of the detailed structure of the theory.
It would be interesting problem to look for holographic dual of the
gauge theories on the squashed sphere,
and confirm that the same result is reproduced by the analysis
on the gravity side.

\section*{Acknowledgements}
We would like to thank K.~Hosomichi and S.~Yokoyama for valuable discussions and comments.
D.~Y. acknowledges the financial support from the Global
Center of Excellence Program by MEXT, Japan through the
``Nanoscience and Quantum Physics'' Project of the Tokyo
Institute of Technology.

\appendix
\section{Separation of long-range and short-range potentials}\label{asym.sec}
In this appendix we determine the explicit form of the long-range and
the short-range potentials.

\begin{figure}[htb]
\centerline{\includegraphics{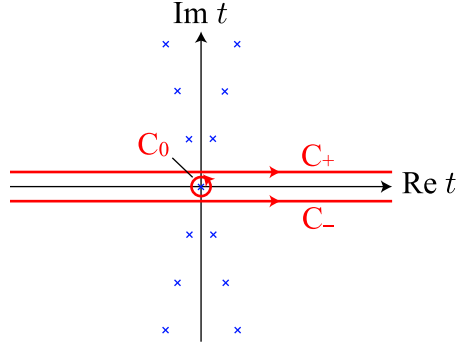}}
\caption{Integration contours $C_\pm$ and $C_0$ on the $t$-plane are shown.
The crosses are poles of function $F(z,t)$.}
\label{contour.eps}
\end{figure}

Let $x$ and $y$ be the real and imaginary parts of $z$.
Namely,
\begin{equation}
z=x+iy.
\end{equation}
In the region $|y|<1/v$,
the function $f_b(z)$ is given by\cite{Kharchev:2001rs,Bytsko:2006ut}
\begin{align}
f_b(z)\equiv \log s_b(z)
&= i\pi\left(\frac{z^2}{2}+\frac{b^2+b^{-2}}{24}\right)
+\int_{C_-}F(z,t)dt
\nonumber\\
&=-i\pi\left(\frac{z^2}{2}+\frac{b^2+b^{-2}}{24}\right)
+\int_{C_+}F(z,t)dt,
\label{fbinteq}
\end{align}
where the function $F(z,t)$ is
\begin{align}
F(z,t) &= \frac{e^{-2itz}}{4t\sinh bt\sinh\frac{t}{b}}
\nonumber\\
&= \frac{1}{4t^3}
-\frac{iz}{2t^2}
-\frac{1}{t}\left(\frac{z^2}{2}+\frac{b^2+b^{-2}}{24}\right)
+\frac{i}{3}z^3+\frac{iz}{12}(b^2+b^{-2})
+{\cal O}(t).
\end{align}
The function $F(z,t)$ has poles at $t=n\pi i b$ and $t=n\pi i b^{-1}$ ($n\in{\bf Z}$).
$C_\pm$ are the contours shown in Fig \ref{contour.eps}.
The first and the second expressions
in (\ref{fbinteq})
are useful for $x>0$ and $x<0$, respectively,
because when $x\rightarrow+\infty$ the integral
in (\ref{fbinteq}) along $C_-$ vanishes,
and when $x\rightarrow+\infty$ the integral along $C_+$ vanishes.
From this fact, we obtain the asymptotic form
\begin{equation}
f_b^{\rm asym}(z)=
i\pi\left(\frac{z^2}{2}+\frac{b^2+b^{-2}}{24}\right)\sign(x).
\label{fasym}
\end{equation}
The difference of $f_b(z)$ from the asymptotic form is
\begin{equation}
f_b(z)-f_b^{\rm asym}(z)
=\int_{C}F(z,t)dt
=\int_{C}F(iy,t)e^{-2itx}dt,
\end{equation}
where $C=C_-$ for $x>0$ and $C=C_+$ for $x<0$.
Because this almost vanishes when $|x|$ is large,
we can approximately express this difference by using $\delta(x)$ as
\begin{equation}
f_b(z)-f_b^{\rm asym}(z)\sim \delta(x)g_b(y)
\end{equation}
We can determine the function $g_b(y)$ by integrating the right hand side
over $x$.
\begin{align}
g_b(y) &= \int_{-\infty}^\infty (f_b(z)-f_b^{\rm asym}(z))dx
\nonumber\\
&=
\int_0^\infty\left(\int_{C_-}F(iy,t)e^{-2itx}dt\right)dx
+\int_{-\infty}^0\left(\int_{C_+}F(iy,t)e^{-2itx}dt\right)dx.
\end{align}
Thanks to small imaginary part of $t$ along the contours $C_\pm$, these $x$ integrals
converge, and we obtain
\begin{align}
g_b(y)
&=
\frac{1}{2i}\int_{C_-}\frac{F(iy,t)}{t}dt
-\frac{1}{2i}\int_{C_+}\frac{F(iy,t)}{t}dt
\nonumber\\
&=
\frac{1}{2i}\oint_{C_0}\frac{F(iy,t)}{t}dt
\nonumber\\
&=
\frac{\pi}{3}y^3-\frac{\pi}{12}(b^2+b^{-2})y.
\label{funcg}
\end{align}


\begin{thebibliography}{99}


\bibitem{Bhattacharya:2008zy}
  J.~Bhattacharya, S.~Bhattacharyya, S.~Minwalla and S.~Raju,
  ``Indices for Superconformal Field Theories in 3,5 and 6 Dimensions,''
  JHEP {\bf 0802}, 064 (2008)
  [arXiv:0801.1435 [hep-th]].
\bibitem{Kim:2009wb}
  S.~Kim,
  ``The complete superconformal index for N=6 Chern-Simons theory,''
  Nucl.\ Phys.\  B {\bf 821}, 241 (2009)
  [arXiv:0903.4172 [hep-th]].
\bibitem{Imamura:2011su}
  Y.~Imamura and S.~Yokoyama,
  ``Index for three dimensional superconformal field theories with general
  R-charge assignments,''
  JHEP {\bf 1104}, 007 (2011)
  [arXiv:1101.0557 [hep-th]].
\bibitem{Kapustin:2009kz}
  A.~Kapustin, B.~Willett, I.~Yaakov,
  ``Exact Results for Wilson Loops in Superconformal Chern-Simons Theories with Matter,''
  JHEP {\bf 1003}, 089 (2010).
  [arXiv:0909.4559 [hep-th]].
\bibitem{Jafferis:2010un}
  D.~L.~Jafferis,
  ``The Exact Superconformal R-Symmetry Extremizes Z,''
  arXiv:1012.3210 [hep-th].
\bibitem{Hama:2010av}
  N.~Hama, K.~Hosomichi and S.~Lee,
  ``Notes on SUSY Gauge Theories on Three-Sphere,''
  JHEP {\bf 1103}, 127 (2011)
  [arXiv:1012.3512 [hep-th]].
\bibitem{Kapustin:2010mh}
  A.~Kapustin, B.~Willett, I.~Yaakov,
  ``Tests of Seiberg-like Duality in Three Dimensions,''
  [arXiv:1012.4021 [hep-th]].
\bibitem{Kapustin:2011gh}
  A.~Kapustin,
  ``Seiberg-like duality in three dimensions for orthogonal gauge groups,''
  arXiv:1104.0466 [hep-th].
\bibitem{Willett:2011gp}
  B.~Willett and I.~Yaakov,
  ``N=2 Dualities and Z Extremization in Three Dimensions,''
  arXiv:1104.0487 [hep-th].
\bibitem{Dolan:2011rp}
  F.~A.~H.~Dolan, V.~P.~Spiridonov, G.~S.~Vartanov,
  ``From 4d superconformal indices to 3d partition functions,''
  [arXiv:1104.1787 [hep-th]].
\bibitem{Jafferis:2011ns}
  D.~Jafferis, X.~Yin,
  ``A Duality Appetizer,''
  [arXiv:1103.5700 [hep-th]].
\bibitem{Benini:2011mf}
  F.~Benini, C.~Closset, S.~Cremonesi,
  ``Comments on 3d Seiberg-like dualities,''
  [arXiv:1108.5373 [hep-th]].
\bibitem{Hwang:2011ht}
  C.~Hwang, K.~-J.~Park, J.~Park,
  ``Evidences for Aharony duality for orthogonal gauge groups,''
  [arXiv:1109.2828 [hep-th]].


\bibitem{Drukker:2010nc}
  N.~Drukker, M.~Marino, P.~Putrov,
  ``From weak to strong coupling in ABJM theory,''
  Commun.\ Math.\ Phys.\  {\bf 306}, 511-563 (2011).
  [arXiv:1007.3837 [hep-th]].
\bibitem{Herzog:2010hf}
  C.~P.~Herzog, I.~R.~Klebanov, S.~S.~Pufu and T.~Tesileanu,
  ``Multi-Matrix Models and Tri-Sasaki Einstein Spaces,''
  Phys.\ Rev.\  D {\bf 83}, 046001 (2011)
  [arXiv:1011.5487 [hep-th]].



\bibitem{Martelli:2011qj}
  D.~Martelli and J.~Sparks,
  ``The large N limit of quiver matrix models and Sasaki-Einstein manifolds,''
  arXiv:1102.5289 [hep-th].
\bibitem{Cheon:2011vi}
  S.~Cheon, H.~Kim and N.~Kim,
  ``Calculating the partition function of N=2 Gauge theories on $S^3$ and
  AdS/CFT correspondence,''
  arXiv:1102.5565 [hep-th].
\bibitem{Jafferis:2011zi}
  D.~L.~Jafferis, I.~R.~Klebanov, S.~S.~Pufu, B.~R.~Safdi,
  ``Towards the F-Theorem: N=2 Field Theories on the Three-Sphere,''
  JHEP {\bf 1106}, 102 (2011).
  [arXiv:1103.1181 [hep-th]].
\bibitem{Imamura:2011uj}
  Y.~Imamura, D.~Yokoyama and S.~Yokoyama,
  ``Superconformal index for large N quiver Chern-Simons theories,''
  JHEP {\bf 1108}, 011 (2011).
  [arXiv:1102.0621 [hep-th]].
\bibitem{Cheon:2011th}
  S.~Cheon, D.~Gang, S.~Kim and J.~Park,
  ``Refined test of AdS4/CFT3 correspondence for N=2,3 theories,''
  arXiv:1102.4273 [hep-th].
\bibitem{Krattenthaler:2011da}
  C.~Krattenthaler, V.~P.~Spiridonov and G.~S.~Vartanov,
  ``Superconformal indices of three-dimensional theories related by mirror
  symmetry,''
  arXiv:1103.4075 [hep-th].
\bibitem{Hwang:2011qt}
  C.~Hwang, H.~Kim, K.~-J.~Park, J.~Park,
  ``Index computation for 3d Chern-Simons matter theory: test of Seiberg-like duality,''
  JHEP {\bf 1109}, 037 (2011).
  [arXiv:1107.4942 [hep-th]].


\bibitem{Hama:2011ea}
  N.~Hama, K.~Hosomichi and S.~Lee,
  ``SUSY Gauge Theories on Squashed Three-Spheres,''
  JHEP {\bf 1105}, 014 (2011).
  [arXiv:1102.4716 [hep-th]].
\bibitem{Gadde:2011ia}
  A.~Gadde and W.~Yan,
  ``Reducing the 4d Index to the $S^3$ Partition Function,''
  arXiv:1104.2592 [hep-th].
\bibitem{Imamura:2011uw}
  Y.~Imamura,
  ``Relation between the 4d superconformal index and the S$^3$ partition function,''
  JHEP {\bf 1109}, 133 (2011).
  [arXiv:1104.4482 [hep-th]].






\bibitem{Kharchev:2001rs}
  S.~Kharchev, D.~Lebedev, M.~Semenov-Tian-Shansky,
  ``Unitary representations of U(q) (sl(2, R)), the modular double, and the multiparticle q deformed Toda chains,''
  Commun.\ Math.\ Phys.\  {\bf 225}, 573-609 (2002).
  [hep-th/0102180].

\bibitem{Bytsko:2006ut}
  A.~G.~Bytsko, J.~Teschner,
  ``Quantization of models with non-compact quantum group symmetry: Modular XXZ magnet and lattice sinh-Gordon model,''
  J.\ Phys.\ A {\bf A39}, 12927-12981 (2006).
  [hep-th/0602093].


\end{thebibliography}
\end{document}